\documentclass[twocolumn,showpacs,superscriptaddress,prl]{revtex4-1}
\usepackage{graphicx}
\usepackage{xcolor}
\usepackage{amsmath,amsthm,amssymb}
\usepackage{braket}
\usepackage{soul}
\usepackage{cancel}
\usepackage[normalem]{ulem} 

\usepackage[colorlinks,citecolor=blue,linkcolor=red]{hyperref}

\begin{document}
 
\title{Coexistence of Dirac Fermions and Magnons in a Layered Two-Dimensional Semiquinoid Metal-Organic Framework}

\author{Christopher Lane}
\email{laneca@lanl.gov}
\affiliation{Theoretical Division, Los Alamos National Laboratory, Los Alamos, New Mexico 87545, USA}
\affiliation{Center for Integrated Nanotechnologies, Los Alamos National Laboratory, Los Alamos, New Mexico 87545, USA}

\author{Yixuan Huang}
\affiliation{Theoretical Division, Los Alamos National Laboratory, Los Alamos, New Mexico 87545, USA}
\affiliation{Center for Integrated Nanotechnologies, Los Alamos National Laboratory, Los Alamos, New Mexico 87545, USA}

\author{Lin Hou}
\affiliation{Theoretical Division, Los Alamos National Laboratory, Los Alamos, New Mexico 87545, USA}

\author{Jian-Xin Zhu}
\affiliation{Theoretical Division, Los Alamos National Laboratory, Los Alamos, New Mexico 87545, USA}
\affiliation{Center for Integrated Nanotechnologies, Los Alamos National Laboratory, Los Alamos, New Mexico 87545, USA}

\date{\today} 
\begin{abstract}
We predict the magnetic and electronic properties of a novel metal-organic framework. By combining density functional theory and density matrix renormalization group approaches, we find the diatomic Kagome crystal structure of the metal-semiquinoid framework (H$_2$NMe$_2$)$_2$M$_2$(Cl$_2$dhbq)$_3$ (M = Ti, V, Cr, Mn, Fe, Co, Ni, Cu, and Zn) to host a rich variety of antiferromagnetic (AFM) and ferromagnetic (FM) Dirac semimetallic, spin-polarized and unpolarized Dirac fermions, along with metallic phases and flat band magnetic insulators. Concomitantly, the spin excitation spectrum of the various magnetic systems display multiple Dirac-like and nodal-ring crossings. This suggests that the metal-semiquinoid system is an ideal platform for examining the intertwining of Dirac fermions and magnons.
\end{abstract}


\maketitle 

\section{Introduction}
Low-dimensional magnetism lies at the heart of numerous novel phenomena in condensed matter physics, including high-$T_{c}$ superconductivity~\cite{fradkin2015colloquium}, topological quantum spin liquids~\cite{zhou2017quantum}, non-Fermi liquid behavior~\cite{stewart2001non}, and quantum criticality~\cite{sachdev2008quantum}. In simple isotropic spin systems with short-range exchange interactions, long-range magnetic order is predicted to be forbidden in one- and two-dimensions due to strong thermal fluctuations~\cite{hohenberg1967existence,mermin1966absence}. However, in real crystalline materials, the spherical atomic symmetry of the magnetic ions is reduced by the local chemical environment. This suggests a straightforward tuning of the magnetic anisotropy can reduce or strengthen spin fluctuations and thus control various forms of magnetic order. Accordingly, the central task in characterizing and designing quantum magnets becomes understanding and manipulating the delicate interplay between spin and orbital motions of an electron, and their coupling to the lattice.

The recent discovery of two-dimensional (2D) van der Waals magnets presents a unique opportunity to study the fundamental processes governing the emergence of magnetic correlations and ordering under 2D confinement, and advance application in spintronics and quantum information sciences~\cite{zhu2021topological,gong2017discovery,huang2017layer,kim2018large,kim2018charge,bonilla2018strong,li2021electronic}. Due to their reduced dimensionality, 2D magnets are an ideal play ground for observing topological superconductivity~\cite{li2021electronic} and frustrated quantum magnetism, such as resonating valence bonds~\cite{anderson1973resonating} and 2D Kitaev spin liquids~\cite{kitaev2006anyons}, which gives us access to fractionalized charge states of matter and new elementary excitations such as Majorana fermions~\cite{huang2022topological,balents2010spin,savary2016quantum,knolle2019field}. However, to date these exotic states have yet to be observed, despite a number of theoretical predictions~\cite{xu2020possible,ji2021rare,loidl2021proximate,takagi2019concept}

Crucial to the discovery of novel magnetic states and the design of next generation miscroelectronics is the control and manipulation of the magnetic properties. To this end, a number of approaches have been fruitful in coupling key parameters governing the underlying electronic structure to external perturbations such as strain, light, gating, proximity and moir\'e heterostructuring~\cite{burch2018magnetism}. For example, interlayer interactions in Fe$_3$GeTe$_2$ thin-films drive a crossover from a single FM domain to an inhomogenous labyrinthine domain phase with increased thickness. This transition is accompanied by a fast increase in the Curie temperature from 130 K to 207 K for just five layers~\cite{fei2018two}. Similarly, when two monolayer sheets of CrI$_3$ are twisted with respect to one another, the resulting moir\'e pattern promotes islands of AFM and FM interlayer coupled domains following the moir\'e pattern~\cite{xu2022coexisting}. Finally, by electrostatic doping and optical pumping, the magnetization, Curie temperature, and polarization of the magnetic state can be directly controlled and fine-tuned~\cite{jiang2018controlling,afanasiev2021controlling}. Despite these significant advances, many of these external perturbations are hard to implement within a realistic environment. Therefore, to be able to incorporate 2D magnetic materials into microcircuitry, we must design the {\it intrinsic} properties of the magnetic thin-films themselves.

Metal-organic frameworks (MOFs) offer a new route to overcome these challenges to build new magnetic materials with high synthetic programmability and tunability. MOFs consist of metal ions coordinated to organic `linker' ligands to form one-, two-, or three-dimensional structures. So far, significant attention has been concentrated on the porosity and chemical stability of MOFs in connection with applications in gas separation and storage, and  catalysis~\cite{wang2019state}. However, recently MOFs have been recognized as an ideal platform materials with desired magnetic and electronic properties~\cite{yamada2017designing,murphy2021exchange}. 2D MOFs in particular, have been proposed~\cite{jiang2021exotic} to host a rich variety of electronic band structure, e.g. Dirac cones and topological flat-bands, and a range of magnetic orders. Despite these intriguing properties, a limited number of compounds with long-range order have been predicted and synthesized.

In this Article, we predict the magnetic and electronic properties of the metal-semiquinoid framework (H$_2$NMe$_2$)$_2$M$_2$(Cl$_2$dhbq)$_3$ (dhbq$^{n-}$ = deprotonated 2,5-dihydroxybenzoquinone) for all $3d$ transition-metals (M = Ti, V, Cr, Mn, Fe, Co, Ni, Cu, and Zn). We find this metal-semiquinoid framework to display a diverse array of magnetic and electronic properties under different transition-metal substitutions. Specifically, long range FM, N\'eel, and Zig-zag AFM magnetic orders are found to be robust within the diatomic Kagome crystal structure. Additionally, the crystal structure in combination with the magnetic state yields a variety of band topologies including AFM and FM Dirac semimetals, spin-polarized Dirac fermions, and flat band magnetic insulators and metals. We calculate magnetic excitation spectrum of each system and find it clearly exhibit multiple Dirac crossings in the FM and zig-zag AFM compounds. Furthermore, multiple compounds studied display the coexistence of Dirac fermions and magnons.

\begin{figure*}[t!]
\includegraphics[width=.99\textwidth]{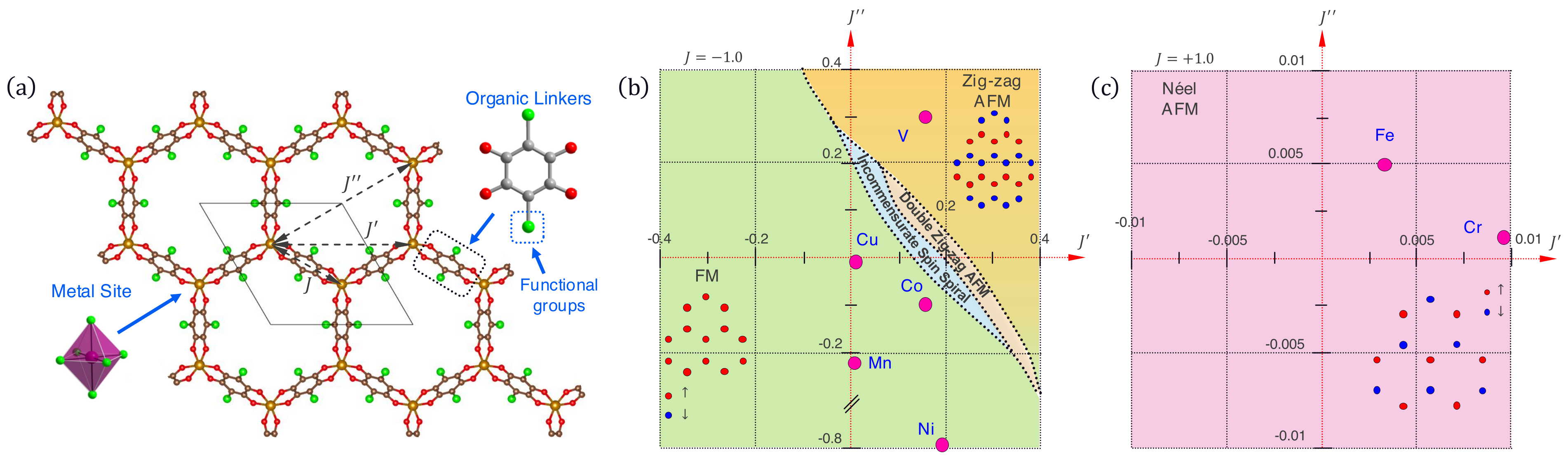}
\caption{(color online) (a) Crystal structure of the metal-semiquinoid framework (H$_2$NMe$_2$)$_2$M$_2$(Cl$_2$dhbq)$_3$, with the transition-metal, oxygen, carbon, and chlorine shown as gold, red, grey, and green spheres. The first-nearest $(J)$, second-nearest $(J^{\prime})$, and third-nearest $(J^{\prime\prime})$ neighbor exchange parameters are indicated on the honeycomb lattice of transition-metal sites within a $2\times 2$ super cell of the metal-semiquinoid framework. The black lines mark the unit cell. (b) and (c) shows the normalized $J$, $J^{\prime}$, and $J^{\prime\prime}$ exchange parameters for all 3$d$ translation metals with $J<0$ and $J>0$, respectively, overlaid on the phase diagram. }
\label{fig:mag_exchange_values}
\end{figure*}

\section{Methods}
{\it Ab initio} calculations were carried out using the pseudopotential projector-augmented wave method~\cite{Kresse1999} implemented in the Vienna ab initio simulation package (VASP)~\cite{Kresse1996,Kresse1993} with an energy cutoff of $400$ eV for the plane-wave basis set. Exchange-correlation effects were treated using the Perdew-Burke-Ernzerhof (PBE) GGA density functional~\cite{perdew1996generalized}. A 7 $\times$ 7 $\times$ 11 $\Gamma$-centered k-point mesh was used to sample the Brillouin zone of the primitive cell. The MOF structure as determined by Jeon {\it et al.}~\cite{jeon20152d} was used in total energy calculations and in initializing the structural relaxation. For the relaxed data, all atomic sites in the unit cell along with the unit cell dimensions were relaxed simultaneously using a conjugate gradient algorithm to minimize energy with an atomic force tolerance of 0.01 eV/\AA. All relaxation calculations included the inert solvents, whereas in the total energy calculations they were removed. A total energy tolerance of $10^{-6}$ eV was used as the convergence criteria on the self-consistent charge density in all cases. 

The magnetic ground state of the effective Heisenberg Hamiltonian was determined by applying the infinite density matrix renormalization group (DMRG)~\cite{white1992density,white1993density,schollwock2011density} approach with $U(1)$ spin symmetry on cylinders with open boundary condition along the axial direction and periodic boundary conditions for the circumferential direction. Results on cylinders of width 6 and 8 unit cells are found to be consistent for each transition metal, where up to 5000 bond dimensions were used to obtain a truncation error of $10^{-5}$. The truncation error is defined as the sum of the discarded reduced density matrix eigenvalues during sweeps. Simulations were performed using the TeNPy library (version 0.9.0)~\cite{hauschild2018efficient}.

\section{Results and Discussion}
\subsection{Magnetic and Electronic Ground State}
Figure~\ref{fig:mag_exchange_values} (a) shows the primitive unit cell of the [M$_{2}$L$_{3}$]$^{2-}$ MOF along the $c$-axis, where the \emph{M}O$_{6}$ octahedra form a 2D honeycomb lattice and the semiquinoid linkers connect the edges of the nearest-neighbor octahedra. In order to determine the exchange coupling strength between neighboring transition-metals, we map the total energies of the various spin configurations $\gamma$ onto those of the third-nearest-neighbor Heisenberg Hamiltonian~\cite{xiang2013magnetic}, written as
\begin{equation}
\mathcal{H}=\sum_{\braket{i<j}}J_{ij}\mathbf{S}_{i}\cdot \mathbf{S}_{j}+\sum_{\braket{\braket{i<j}}}J_{ij}^{\prime}\mathbf{S}_{i}\cdot \mathbf{S}_{j}+\sum_{\braket{\braket{\braket{i<j}}}}J_{ij}^{\prime\prime}\mathbf{S}_{i}\cdot \mathbf{S}_{j}.
\label{eq:Hnn}
\end{equation}
Here $i(j)$ indexes the magnetic ion lattice positions, $S_{i}$ is the local magnetic moment on lattice site $i$. $J_{ij}$, $J_{ij}^{\prime}$, and $J_{ij}^{\prime\prime}$ denote the nearest-neighbor, next-nearest-neighbor, and third-nearest-neighbor exchange interaction strength, with the symmetry properties of $J_{ij}^{(\prime,\prime\prime)}=J_{ji}^{(\prime,\prime\prime)}$. Assuming a fully isotropic bond independent model, $2^{4}$ unique spin configurations are needed to fully determine $J^{\prime\cdots \prime}$. By summing configurations $\gamma$ and $-\gamma$, and using the resulting eight equations to solve for $J$, $J^{\prime}$, and $J^{\prime\prime}$, we find 
{\small
\begin{subequations}
\begin{align}
J=\frac{(E^1+E^4+E^5+E^6)-(E^2+E^3+E^7+E^8)}{16\cdot 3\braket{S}^{2}},\\
J^{\prime}=\frac{(E^1+E^3+E^5+E^7)-(E^2+E^4+E^6+E^8)}{16\cdot 6\braket{S}^{2}},\\
J^{\prime\prime}=\frac{(E^1+E^3+E^4+E^8)-(E^2+E^5+E^6+E^7)}{16\cdot 3\braket{S}^{2}},
\end{align}
\end{subequations}
}
where,
\begin{align}
E^{1}=E^{++++}+E^{----}, ~&~ E^{5}=E^{+++-}+E^{---+},\nonumber\\
E^{2}=E^{-+++}+E^{+---}, ~&~ E^{6}=E^{++--}+E^{--++},\nonumber\\
E^{3}=E^{+-++}+E^{-+--}, ~&~ E^{7}=E^{+-+-}+E^{-+-+},\nonumber\\
E^{4}=E^{++-+}+E^{--+-}, ~&~ E^{8}=E^{+--+}+E^{-++-}.\nonumber
\end{align}
Here, each spin configuration $\gamma$ is given by the direction of the magnetic moment on each successive nearest-neighbor ring, e.g., $+-+-$. In calculating the total energies, we found $E^{\gamma} \approx E^{-\gamma}$, thereby allowing us to reduce the number of {\it ab intio} total energy calculations by half. Moreover, the H$_2$NMe$_2$ and H$_{2}$O solvent molecules have been neglected for simplicity in the band dispersion and total energy calculations.

Figure~\ref{fig:mag_exchange_values} (b) presents the normalized first-, second-, and third-nearest neighbor exchange parameters for all 3$d$ transition metals in the [M$_{2}$L$_{3}$]$^{2-}$ MOF. Ti and Zn display negligible and zero magnetic polarization on the atomic sites, respectively, yielding no exchange parameters. For transition-metals to the left of Co on the periodic table, $J$ and $J^{\prime\prime}$ alternate between positive and negative values, whereas Co, Ni, and Cu are consistently negative. The next-nearest neighbor exchange coupling is found to be finite, despite the 15.66~\AA~ separation, and positive for all magnetic ions. Interestingly, this oscillatory behavior in the sign of the exchange coupling indicates a possible sensitive dependence of the ground state magnetic order on the $d$-electron count. For the value of the exchange parameters and corresponding magnetic moment, see Table~S1 in the Supplementary Materials~\cite{SuppMaterial}.

The magnetic ground state of [M$_{2}$L$_{3}$]$^{2-}$ MOF for each transition metal is determined by solving for the ground state of the next-next-nearest neighbor Heisenberg model [Eq.~(\ref{eq:Hnn})] within the DMRG. The resulting magnetic ground state for each transition metal is indicated in Fig.~\ref{fig:mag_exchange_values}(b) and (c). Note the indicated phase boundaries are for the various transition-metals with $\braket{S}=1$ and does not facilitate direct comparison between species since $\braket{S}$ is not necessarily equivalent.
We find every transition-metal ion to be in the {\it high} spin state, with spin quantum numbers $S = 1/2$ for Cu, $S = 1$ for V and Ni, $S = 3/2$ for Cr, Mn, and Co, and $S = 2$ for Fe. Specifically, vanadium exhibits a zig-zag AFM ground state in close proximity to a double zig-zag, incommensurate spin spiral, and FM phase transition. The zig-zag AFM, FM~\cite{watanabe2022frustrated, bose2022proximate,jiang2023quantum}, and double zig-zag~\cite{jiang2023quantum} phases are also found in similar studies for $\braket{S}=1/2$, but the incommensurate spin spiral phase is absent due to enhanced quantum fluctuations for smaller $\braket{S}$. As the atomic number increases, the ground state alternates between FM and N\'{e}el-type AFM order up through Fe. For Co, Ni, and Cu the ground state settles into a robust FM order. Similar N\'{e}el-type AFM and FM phases are also found in other 2D magnet families including Cr$_2$(Si,Ge)$_2$Te$_6$~\cite{gong2017discovery}, and MnPS$_3$~\cite{wildes1994true}.

\begin{figure*}[ht!]
\includegraphics[width=1\textwidth]{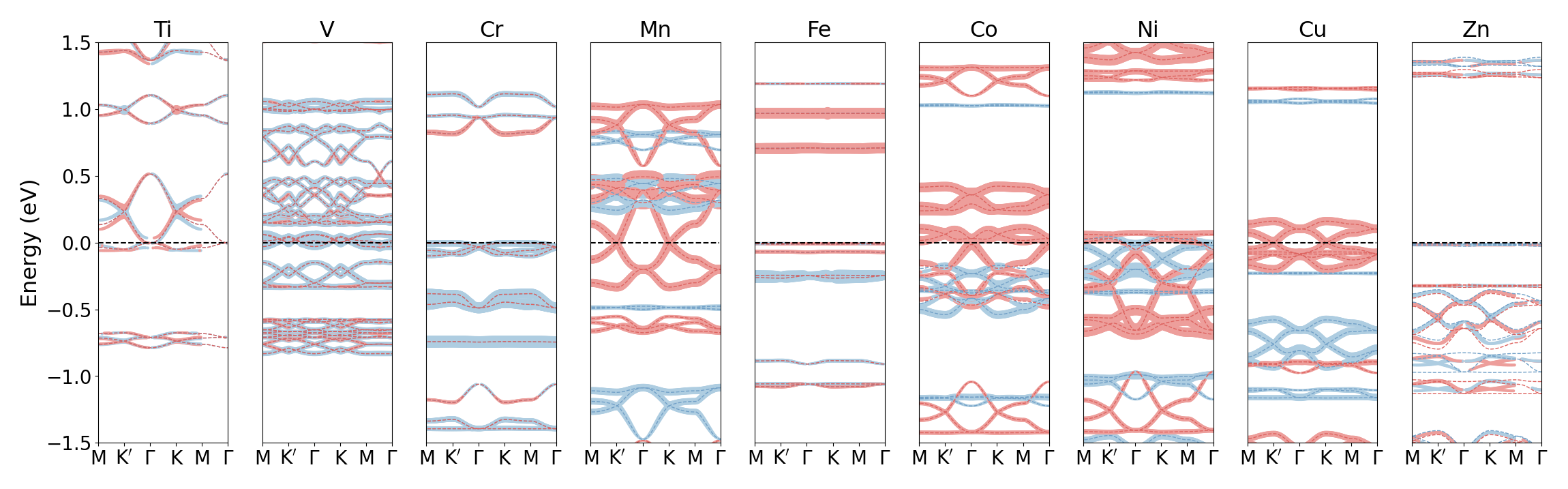}
\caption{(color online) Spin-resolved electronic band dispersions with (dots) and without (dashed line) spin-orbit coupling of the metal-semiquinoid framework (H$_2$NMe$_2$)$_2$M$_2$(Cl$_2$dhbq)$_3$ for various $3d$ transition-metals. The sizes of the dots are proportional to the spin-polarization $\braket{S_{z}}$, with the spin-up (down) states indicated in red (blue) color. } 
\label{fig:bands}
\end{figure*}

Presently, there are no experimental studies investigating the magnetic properties of these 2D MOFs besides the Fe variant. Here, we compare our results to available experimental data on [Fe$_{2}$L$_{3}$]$^{2-}$. We predict a value of the magnetic moment on iron sites of 3.922$\mu_B$, which is consistent with an octahedrally coordinated High-spin Fe$^{+3}$ ion. This is in excellent accord with M\"{o}ssbauer spectra indicating a high-spin Fe$^{+3}$ ion, as reported by Jeon {\it et al.}~\cite{jeon20152d}.

To understand the sensitivity of the magnetic ground state to structural changes which are prevalent in MOFs due to their flexible and floppy nature, we performed additional DFT and DMRG calculations on the relaxed MOF crystal structure. The H$_2$NMe$_2$ and H$_{2}$O solvent molecules were included in the crystal relaxation. Here, the magnetic moment on the iron sites is 1.038$\mu_B$, which is congruent with the low-spin state of octahedrally corrdinated Fe$^{+3}$ ions. Interestingly, the slightly reduced unit cell (by $\sim$ 19.6\%) of the relaxed structure yields the low-spin state of Fe$^{+3}$, thereby suggesting a subtle competition between crystal field splitting and Hund's coupling energy scales in the [M$_{2}$L$_{3}$]$^{2-}$ MOF. Furthermore, upon performing these calculations for all 3d transition-metals reveals the same trend. That is, all transition-metal ions in the relaxed structures display a low-spin state, whereas the experimental crystal structures display the high-spin state. See Sec.~S2 of the Supplementary Materials~\cite{SuppMaterial} for a summary of the theoretical results for the relaxed crystal structure. Furthermore, this indicated that the experimental crystal structure captures static entropic effects that are present at finite temperatures that are necessary to accurately module this MOF system. Overall, this suggests the magnetism in the [M$_{2}$L$_{3}$]$^{2-}$ MOF sensitively depends on the theoretical approach (unrelaxed/relaxed). Moreover, this suggests future {\it ab initio} molecular dynamics simulation should yield insights in the temperature dependence of the interplay between magnetism, crystal-field splitting, and Hund's electronic interactions.  

\begin{figure*}[t!]
\includegraphics[width=.99\textwidth]{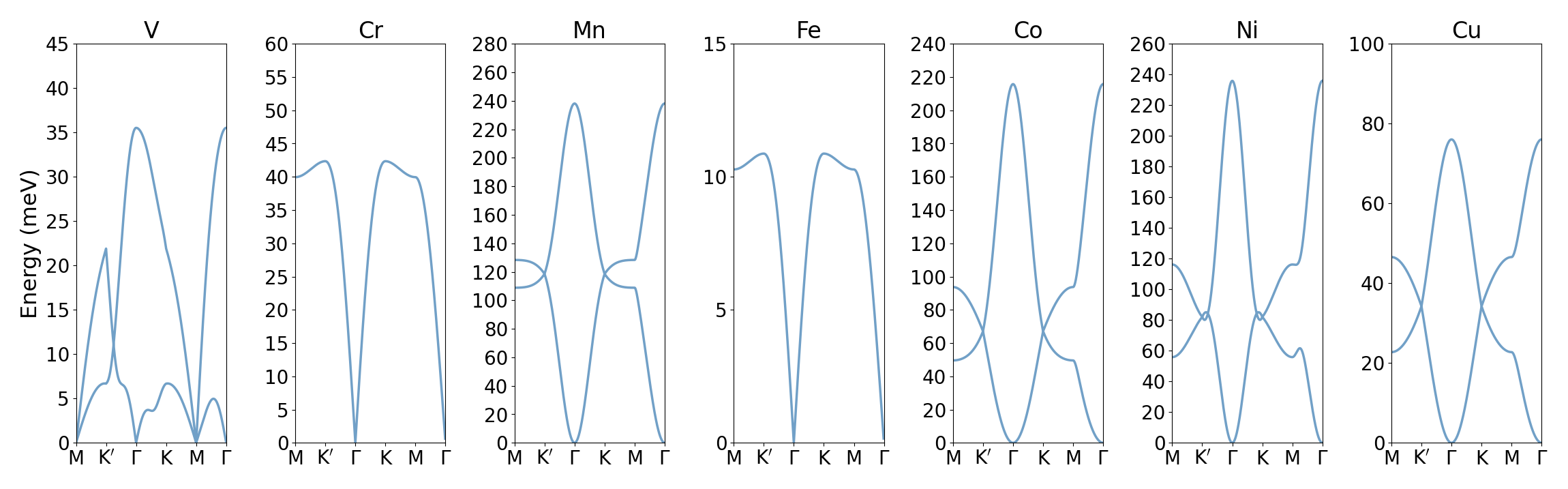}
\caption{(color online) Magnon excitation spectrum of the metal-semiquinoid framework (H$_2$NMe$_2$)$_2$M$_2$(Cl$_2$dhbq)$_3$ for various $3d$ transition-metals.} 
\label{fig:magnon}
\end{figure*}

Structurally, [M$_{2}$L$_{3}$]$^{2-}$ may be mapped onto the diatomic kagome lattice resulting from the $\sim 90^{\circ}$ dihedral angles between the organic ligands~\cite{ni2020pi}. This geometry is predicted to produce two sets of enantimorphic Yin-Yang kagome bands with a Dirac crossing at $K(K^\prime)$ in the Brillouin zone. When spin-orbit coupling is included a finite non-trivial gap appears at the Dirac nodes, thereby transforming the system from a topological semimetal to a topological insulator with two flat bands of opposite spin Chern numbers~\cite{jiang2021exotic}. Such diatomic kagome crystals have been proposed as a platform for a triplet excitonic insulator state~\cite{sethi2021flat} and to host novel quantum Hall effects~\cite{zhou2019excited}.

Figure~\ref{fig:bands} shows the spin-resolved electronic band dispersion for each [M$_{2}$L$_{3}$]$^{2-}$ in its associated magnetic ground state. Compared to idealized tight-binding models~\cite{jiang2021exotic}, a rich variety of electronic structures are exhibited showing a intertwining of electronic band properties, magnetic ground state, and $d$-electron count. Specifically, Cr, Fe, and Zn display an AFM and a NM insulating state, respectively, with highly localized carriers as demonstrated by the extremely flat bands at the valence band edges, similar to those in kagome and  magic-angle twisted bilayer graphene~\cite{cao2018correlated}. 

Mn and Cu host fully spin-polarized Dirac fermions at the Fermi level that are clearly separated from other bands in the system by $\sim 250$ meV. Interestingly, the bandwidth and Fermi velocity of the linearly dispersing states of the Cu-based compound are significantly reduced compared to those in Mn, indicating an enhancement in the electron–electron interaction strength. Many theoretical studies suggest that Dirac materials with strong electron–electron interactions may host an excitonic insulator ground state~\cite{pertsova2020dynamically,kotov2012electron,drut2009graphene}, however, such a phase has yet to be observed in a Dirac material. In contrast, FM Ni is a metal and FM Co is a narrow in-direct band gap ($\sim 25$ meV) FM semiconductor. Finally, under electron doping V and Ti are in close proximity to a AFM Dirac semimetal and the elusive nearly nonmagnetic spin–valley coupled Dirac semimetal phase, respectively.

\subsection{Magnon Excitation Spectrum}
We now consider the elementary excitations of the FM, N\'eel AFM and zig-zag AFM phases. To derive their dispersion relations, we map the spin operators in the Heisenberg model, Eq.~(\ref{eq:Hnn}), onto a set of boson creation (annihilation) operators through the Holstein-Primakoff transformation~\cite{holstein1940field,khomskii2010basic}. By retaining the lowest-order terms, the dynamical matrix is diagonalized by employing a Bogoliubov canonical transformation~\footnote{We note that in contrast to the Fermi systems where the dynamical excitation matrix is identical to the Hamiltonian coefficient matrix, the dynamic matrix is distinct in the case of Bose systems~\cite{xiao2009theory}. Therefore care must be taken when calculating the magnon excitation spectrum.} (see the Supplementary Materials for details~\cite{SuppMaterial}), which yields the magnon energy dispersions
\begin{subequations}
\begin{align}
&\varepsilon^{\pm}_{\text{FM}}(\mathbf{k})=S(\chi_{\text{FM}}(\mathbf{k})\pm |\phi (\mathbf{k})|), \\
&\varepsilon_{\text{N\'eel}}(\mathbf{k})=S\sqrt{\chi_{\text{N\'eel}}^{2}(\mathbf{k})- |\phi (\mathbf{k})|^{2}},\\
&\varepsilon^{\pm}_{\text{zig-zag}}(\mathbf{k})=S\left(\chi_{1}^{2}+\left|\phi_{1}\right|^{2}-\left|\phi_{2}\right|^{2}-\frac{1}{2}\left|\phi_{3}\right|^{2}-\frac{1}{2}\left|\phi_{4}\right|^{2}\right. \nonumber\\ 
&\left.\pm \sqrt{F(\chi_{1},\phi_{1},\phi_{2},\phi_{3},\phi_{4})}\right)^{1/2}
\end{align}
\end{subequations}
where $\chi_{\text{FM}}(\mathbf{k}),\chi_{\text{N\'eel}}(\mathbf{k}),\phi(\mathbf{k})$ and $\chi_1,\phi_1,\phi_2,\phi_3,\phi_4, F$ are functions of $\mathbf{k}$ and $J^{\prime\cdots\prime}$; see their definition in Sec.~S4 in the Supplementary Materials~\cite{SuppMaterial}.

Figure~\ref{fig:magnon} presents the magnon dispersion for each 3$d$ translation metal. In Cr and Fe based semiquinoid metal-organic frameworks a single magnon band is found that follows a linear dispersion to zero at $\Gamma$, as is typical for antiferromagents. The FM Mn, Co, and Cu based compounds display clear Dirac crossings at K (K$^\prime$), along with the standard quadratic band dispersion at $\Gamma$. Ni exhibits a Dirac magnon nodal loop encircling the K (K$^\prime$) point in the Brillouin zone, similar to that proposed in the honeycomb chromium halide compounds~\cite{owerre2017dirac}. Finally, the the zig-zag AFM state of V produces a very sharp Dirac crossing near K$^\prime$ and two linearly dispersing bands at M due to the finite ordering vector of the underlying magnetic order.

Dirac fermions display a wide range of unique properties including large opacity, huge charge carrier mobility and the anomalous quantum Hall effect~\cite{novoselov2011nobel,neto2009electronic}. Similarly, magnetic systems have been predicted and observed to host Dirac magnons~\cite{fransson2016magnon,owerre2016first,pershoguba2018dirac}, exhibiting gapped bands with non-trivial Chern numbers~\cite{zhu2021topological}, surface states~\cite{pershoguba2018dirac}, and a magnon thermal Hall effect~\cite{onose2010observation}. Here, we find the [M$_{2}$L$_{3}$]$^{2-}$ MOF composed of V, Mn, Co, Ni, and Cu to display the coexistence of Direc fermions and magnons which allows for the intertwining of non-trivial fermionic and bosonic band topologies.  Consequently, by combining dissipationless long distance magnon and charge propagation, this family of materials could provide a path towards charge-magnetic coupled multifunctional magnonic and spintronic devices. 

\section{Conclusions}
We have shown the semiquinoid-based metal-organic framework (H$_2$NMe$_2$)$_2$M$_2$(Cl$_2$dhbq)$_3$ to host a rich landscape of electronic and magnetic properties for $3d$ transition-metals. In particular, Mn demonstrates the coexistence of Dirac fermions and magnons, which hold promise is dissipationless charge and spin transport and unique non-trivial fermion-magnon coupled band topologies. Moreover, the inclusion of higher order corrections, e.g., ring exchange~\cite{coldea2001spin}, can open a magnon gap~\cite{li2022ring} and give rise to possible spin liquid phases~\cite{cookmeyer2021four}.

\section{Acknowledgments}
\begin{acknowledgments}
We thank Jennifer Hollingsworth and Ekaterina Dogopolova for useful discussions.
This work was carried out under the auspices of the U.S. Department of Energy (DOE) National Nuclear Security Administration (NNSA) under Contract No. 89233218CNA000001. It was supported by the LDRD program (C.L.),  Center for Integrated Nanotechnologies (Y.H.), a DOE BES user facility, in partnership with the LANL Institutional Computing Program for computational resources, and Quantum Science Center (J.-X.Z.), a U.S. DOE Office of Science Quantum Information Science Research Center.
\end{acknowledgments}

\bibliography{MOF_Refs}

\begin{thebibliography}{68}%
\makeatletter
\providecommand \@ifxundefined [1]{%
 \@ifx{#1\undefined}
}%
\providecommand \@ifnum [1]{%
 \ifnum #1\expandafter \@firstoftwo
 \else \expandafter \@secondoftwo
 \fi
}%
\providecommand \@ifx [1]{%
 \ifx #1\expandafter \@firstoftwo
 \else \expandafter \@secondoftwo
 \fi
}%
\providecommand \natexlab [1]{#1}%
\providecommand \enquote  [1]{``#1''}%
\providecommand \bibnamefont  [1]{#1}%
\providecommand \bibfnamefont [1]{#1}%
\providecommand \citenamefont [1]{#1}%
\providecommand \href@noop [0]{\@secondoftwo}%
\providecommand \href [0]{\begingroup \@sanitize@url \@href}%
\providecommand \@href[1]{\@@startlink{#1}\@@href}%
\providecommand \@@href[1]{\endgroup#1\@@endlink}%
\providecommand \@sanitize@url [0]{\catcode `\\12\catcode `\$12\catcode
  `\&12\catcode `\#12\catcode `\^12\catcode `\_12\catcode `\%12\relax}%
\providecommand \@@startlink[1]{}%
\providecommand \@@endlink[0]{}%
\providecommand \url  [0]{\begingroup\@sanitize@url \@url }%
\providecommand \@url [1]{\endgroup\@href {#1}{\urlprefix }}%
\providecommand \urlprefix  [0]{URL }%
\providecommand \Eprint [0]{\href }%
\providecommand \doibase [0]{http://dx.doi.org/}%
\providecommand \selectlanguage [0]{\@gobble}%
\providecommand \bibinfo  [0]{\@secondoftwo}%
\providecommand \bibfield  [0]{\@secondoftwo}%
\providecommand \translation [1]{[#1]}%
\providecommand \BibitemOpen [0]{}%
\providecommand \bibitemStop [0]{}%
\providecommand \bibitemNoStop [0]{.\EOS\space}%
\providecommand \EOS [0]{\spacefactor3000\relax}%
\providecommand \BibitemShut  [1]{\csname bibitem#1\endcsname}%
\let\auto@bib@innerbib\@empty
\bibitem [{\citenamefont {Fradkin}\ \emph {et~al.}(2015)\citenamefont
  {Fradkin}, \citenamefont {Kivelson},\ and\ \citenamefont
  {Tranquada}}]{fradkin2015colloquium}%
  \BibitemOpen
  \bibfield  {author} {\bibinfo {author} {\bibfnamefont {E.}~\bibnamefont
  {Fradkin}}, \bibinfo {author} {\bibfnamefont {S.~A.}\ \bibnamefont
  {Kivelson}}, \ and\ \bibinfo {author} {\bibfnamefont {J.~M.}\ \bibnamefont
  {Tranquada}},\ }\href@noop {} {\bibfield  {journal} {\bibinfo  {journal}
  {Reviews of Modern Physics}\ }\textbf {\bibinfo {volume} {87}},\ \bibinfo
  {pages} {457} (\bibinfo {year} {2015})}\BibitemShut {NoStop}%
\bibitem [{\citenamefont {Zhou}\ \emph {et~al.}(2017)\citenamefont {Zhou},
  \citenamefont {Kanoda},\ and\ \citenamefont {Ng}}]{zhou2017quantum}%
  \BibitemOpen
  \bibfield  {author} {\bibinfo {author} {\bibfnamefont {Y.}~\bibnamefont
  {Zhou}}, \bibinfo {author} {\bibfnamefont {K.}~\bibnamefont {Kanoda}}, \ and\
  \bibinfo {author} {\bibfnamefont {T.-K.}\ \bibnamefont {Ng}},\ }\href@noop {}
  {\bibfield  {journal} {\bibinfo  {journal} {Reviews of Modern Physics}\
  }\textbf {\bibinfo {volume} {89}},\ \bibinfo {pages} {025003} (\bibinfo
  {year} {2017})}\BibitemShut {NoStop}%
\bibitem [{\citenamefont {Stewart}(2001)}]{stewart2001non}%
  \BibitemOpen
  \bibfield  {author} {\bibinfo {author} {\bibfnamefont {G.}~\bibnamefont
  {Stewart}},\ }\href@noop {} {\bibfield  {journal} {\bibinfo  {journal}
  {Reviews of Modern Physics}\ }\textbf {\bibinfo {volume} {73}},\ \bibinfo
  {pages} {797} (\bibinfo {year} {2001})}\BibitemShut {NoStop}%
\bibitem [{\citenamefont {Sachdev}(2008)}]{sachdev2008quantum}%
  \BibitemOpen
  \bibfield  {author} {\bibinfo {author} {\bibfnamefont {S.}~\bibnamefont
  {Sachdev}},\ }\href@noop {} {\bibfield  {journal} {\bibinfo  {journal}
  {Nature Physics}\ }\textbf {\bibinfo {volume} {4}},\ \bibinfo {pages} {173}
  (\bibinfo {year} {2008})}\BibitemShut {NoStop}%
\bibitem [{\citenamefont {Hohenberg}(1967)}]{hohenberg1967existence}%
  \BibitemOpen
  \bibfield  {author} {\bibinfo {author} {\bibfnamefont {P.~C.}\ \bibnamefont
  {Hohenberg}},\ }\href@noop {} {\bibfield  {journal} {\bibinfo  {journal}
  {Physical Review}\ }\textbf {\bibinfo {volume} {158}},\ \bibinfo {pages}
  {383} (\bibinfo {year} {1967})}\BibitemShut {NoStop}%
\bibitem [{\citenamefont {Mermin}\ and\ \citenamefont
  {Wagner}(1966)}]{mermin1966absence}%
  \BibitemOpen
  \bibfield  {author} {\bibinfo {author} {\bibfnamefont {N.~D.}\ \bibnamefont
  {Mermin}}\ and\ \bibinfo {author} {\bibfnamefont {H.}~\bibnamefont
  {Wagner}},\ }\href@noop {} {\bibfield  {journal} {\bibinfo  {journal}
  {Physical Review Letters}\ }\textbf {\bibinfo {volume} {17}},\ \bibinfo
  {pages} {1133} (\bibinfo {year} {1966})}\BibitemShut {NoStop}%
\bibitem [{\citenamefont {Zhu}\ \emph {et~al.}(2021)\citenamefont {Zhu},
  \citenamefont {Zhang}, \citenamefont {Wang}, \citenamefont {Dos~Santos},
  \citenamefont {Song}, \citenamefont {Mueller}, \citenamefont {Schmalzl},
  \citenamefont {Schmidt}, \citenamefont {Ivanov}, \citenamefont {Park} \emph
  {et~al.}}]{zhu2021topological}%
  \BibitemOpen
  \bibfield  {author} {\bibinfo {author} {\bibfnamefont {F.}~\bibnamefont
  {Zhu}}, \bibinfo {author} {\bibfnamefont {L.}~\bibnamefont {Zhang}}, \bibinfo
  {author} {\bibfnamefont {X.}~\bibnamefont {Wang}}, \bibinfo {author}
  {\bibfnamefont {F.~J.}\ \bibnamefont {Dos~Santos}}, \bibinfo {author}
  {\bibfnamefont {J.}~\bibnamefont {Song}}, \bibinfo {author} {\bibfnamefont
  {T.}~\bibnamefont {Mueller}}, \bibinfo {author} {\bibfnamefont
  {K.}~\bibnamefont {Schmalzl}}, \bibinfo {author} {\bibfnamefont {W.~F.}\
  \bibnamefont {Schmidt}}, \bibinfo {author} {\bibfnamefont {A.}~\bibnamefont
  {Ivanov}}, \bibinfo {author} {\bibfnamefont {J.~T.}\ \bibnamefont {Park}},
  \emph {et~al.},\ }\href@noop {} {\bibfield  {journal} {\bibinfo  {journal}
  {Science Advances}\ }\textbf {\bibinfo {volume} {7}},\ \bibinfo {pages}
  {eabi7532} (\bibinfo {year} {2021})}\BibitemShut {NoStop}%
\bibitem [{\citenamefont {Gong}\ \emph {et~al.}(2017)\citenamefont {Gong},
  \citenamefont {Li}, \citenamefont {Li}, \citenamefont {Ji}, \citenamefont
  {Stern}, \citenamefont {Xia}, \citenamefont {Cao}, \citenamefont {Bao},
  \citenamefont {Wang}, \citenamefont {Wang} \emph
  {et~al.}}]{gong2017discovery}%
  \BibitemOpen
  \bibfield  {author} {\bibinfo {author} {\bibfnamefont {C.}~\bibnamefont
  {Gong}}, \bibinfo {author} {\bibfnamefont {L.}~\bibnamefont {Li}}, \bibinfo
  {author} {\bibfnamefont {Z.}~\bibnamefont {Li}}, \bibinfo {author}
  {\bibfnamefont {H.}~\bibnamefont {Ji}}, \bibinfo {author} {\bibfnamefont
  {A.}~\bibnamefont {Stern}}, \bibinfo {author} {\bibfnamefont
  {Y.}~\bibnamefont {Xia}}, \bibinfo {author} {\bibfnamefont {T.}~\bibnamefont
  {Cao}}, \bibinfo {author} {\bibfnamefont {W.}~\bibnamefont {Bao}}, \bibinfo
  {author} {\bibfnamefont {C.}~\bibnamefont {Wang}}, \bibinfo {author}
  {\bibfnamefont {Y.}~\bibnamefont {Wang}},  \emph {et~al.},\ }\href@noop {}
  {\bibfield  {journal} {\bibinfo  {journal} {Nature}\ }\textbf {\bibinfo
  {volume} {546}},\ \bibinfo {pages} {265} (\bibinfo {year}
  {2017})}\BibitemShut {NoStop}%
\bibitem [{\citenamefont {Huang}\ \emph {et~al.}(2017)\citenamefont {Huang},
  \citenamefont {Clark}, \citenamefont {Navarro-Moratalla}, \citenamefont
  {Klein}, \citenamefont {Cheng}, \citenamefont {Seyler}, \citenamefont
  {Zhong}, \citenamefont {Schmidgall}, \citenamefont {McGuire}, \citenamefont
  {Cobden} \emph {et~al.}}]{huang2017layer}%
  \BibitemOpen
  \bibfield  {author} {\bibinfo {author} {\bibfnamefont {B.}~\bibnamefont
  {Huang}}, \bibinfo {author} {\bibfnamefont {G.}~\bibnamefont {Clark}},
  \bibinfo {author} {\bibfnamefont {E.}~\bibnamefont {Navarro-Moratalla}},
  \bibinfo {author} {\bibfnamefont {D.~R.}\ \bibnamefont {Klein}}, \bibinfo
  {author} {\bibfnamefont {R.}~\bibnamefont {Cheng}}, \bibinfo {author}
  {\bibfnamefont {K.~L.}\ \bibnamefont {Seyler}}, \bibinfo {author}
  {\bibfnamefont {D.}~\bibnamefont {Zhong}}, \bibinfo {author} {\bibfnamefont
  {E.}~\bibnamefont {Schmidgall}}, \bibinfo {author} {\bibfnamefont {M.~A.}\
  \bibnamefont {McGuire}}, \bibinfo {author} {\bibfnamefont {D.~H.}\
  \bibnamefont {Cobden}},  \emph {et~al.},\ }\href@noop {} {\bibfield
  {journal} {\bibinfo  {journal} {Nature}\ }\textbf {\bibinfo {volume} {546}},\
  \bibinfo {pages} {270} (\bibinfo {year} {2017})}\BibitemShut {NoStop}%
\bibitem [{\citenamefont {Kim}\ \emph {et~al.}(2018{\natexlab{a}})\citenamefont
  {Kim}, \citenamefont {Seo}, \citenamefont {Lee}, \citenamefont {Ko},
  \citenamefont {Kim}, \citenamefont {Jang}, \citenamefont {Ok}, \citenamefont
  {Lee}, \citenamefont {Jo}, \citenamefont {Kang} \emph
  {et~al.}}]{kim2018large}%
  \BibitemOpen
  \bibfield  {author} {\bibinfo {author} {\bibfnamefont {K.}~\bibnamefont
  {Kim}}, \bibinfo {author} {\bibfnamefont {J.}~\bibnamefont {Seo}}, \bibinfo
  {author} {\bibfnamefont {E.}~\bibnamefont {Lee}}, \bibinfo {author}
  {\bibfnamefont {K.-T.}\ \bibnamefont {Ko}}, \bibinfo {author} {\bibfnamefont
  {B.}~\bibnamefont {Kim}}, \bibinfo {author} {\bibfnamefont {B.~G.}\
  \bibnamefont {Jang}}, \bibinfo {author} {\bibfnamefont {J.~M.}\ \bibnamefont
  {Ok}}, \bibinfo {author} {\bibfnamefont {J.}~\bibnamefont {Lee}}, \bibinfo
  {author} {\bibfnamefont {Y.~J.}\ \bibnamefont {Jo}}, \bibinfo {author}
  {\bibfnamefont {W.}~\bibnamefont {Kang}},  \emph {et~al.},\ }\href@noop {}
  {\bibfield  {journal} {\bibinfo  {journal} {Nature Materials}\ }\textbf
  {\bibinfo {volume} {17}},\ \bibinfo {pages} {794} (\bibinfo {year}
  {2018}{\natexlab{a}})}\BibitemShut {NoStop}%
\bibitem [{\citenamefont {Kim}\ \emph {et~al.}(2018{\natexlab{b}})\citenamefont
  {Kim}, \citenamefont {Kim}, \citenamefont {Sandilands}, \citenamefont {Sinn},
  \citenamefont {Lee}, \citenamefont {Son}, \citenamefont {Lee}, \citenamefont
  {Choi}, \citenamefont {Kim}, \citenamefont {Park} \emph
  {et~al.}}]{kim2018charge}%
  \BibitemOpen
  \bibfield  {author} {\bibinfo {author} {\bibfnamefont {S.~Y.}\ \bibnamefont
  {Kim}}, \bibinfo {author} {\bibfnamefont {T.~Y.}\ \bibnamefont {Kim}},
  \bibinfo {author} {\bibfnamefont {L.~J.}\ \bibnamefont {Sandilands}},
  \bibinfo {author} {\bibfnamefont {S.}~\bibnamefont {Sinn}}, \bibinfo {author}
  {\bibfnamefont {M.-C.}\ \bibnamefont {Lee}}, \bibinfo {author} {\bibfnamefont
  {J.}~\bibnamefont {Son}}, \bibinfo {author} {\bibfnamefont {S.}~\bibnamefont
  {Lee}}, \bibinfo {author} {\bibfnamefont {K.-Y.}\ \bibnamefont {Choi}},
  \bibinfo {author} {\bibfnamefont {W.}~\bibnamefont {Kim}}, \bibinfo {author}
  {\bibfnamefont {B.-G.}\ \bibnamefont {Park}},  \emph {et~al.},\ }\href@noop
  {} {\bibfield  {journal} {\bibinfo  {journal} {Physical Review Letters}\
  }\textbf {\bibinfo {volume} {120}},\ \bibinfo {pages} {136402} (\bibinfo
  {year} {2018}{\natexlab{b}})}\BibitemShut {NoStop}%
\bibitem [{\citenamefont {Bonilla}\ \emph {et~al.}(2018)\citenamefont
  {Bonilla}, \citenamefont {Kolekar}, \citenamefont {Ma}, \citenamefont {Diaz},
  \citenamefont {Kalappattil}, \citenamefont {Das}, \citenamefont {Eggers},
  \citenamefont {Gutierrez}, \citenamefont {Phan},\ and\ \citenamefont
  {Batzill}}]{bonilla2018strong}%
  \BibitemOpen
  \bibfield  {author} {\bibinfo {author} {\bibfnamefont {M.}~\bibnamefont
  {Bonilla}}, \bibinfo {author} {\bibfnamefont {S.}~\bibnamefont {Kolekar}},
  \bibinfo {author} {\bibfnamefont {Y.}~\bibnamefont {Ma}}, \bibinfo {author}
  {\bibfnamefont {H.~C.}\ \bibnamefont {Diaz}}, \bibinfo {author}
  {\bibfnamefont {V.}~\bibnamefont {Kalappattil}}, \bibinfo {author}
  {\bibfnamefont {R.}~\bibnamefont {Das}}, \bibinfo {author} {\bibfnamefont
  {T.}~\bibnamefont {Eggers}}, \bibinfo {author} {\bibfnamefont {H.~R.}\
  \bibnamefont {Gutierrez}}, \bibinfo {author} {\bibfnamefont {M.-H.}\
  \bibnamefont {Phan}}, \ and\ \bibinfo {author} {\bibfnamefont
  {M.}~\bibnamefont {Batzill}},\ }\href@noop {} {\bibfield  {journal} {\bibinfo
   {journal} {Nature Nanotechnology}\ }\textbf {\bibinfo {volume} {13}},\
  \bibinfo {pages} {289} (\bibinfo {year} {2018})}\BibitemShut {NoStop}%
\bibitem [{\citenamefont {Li}\ \emph {et~al.}(2021)\citenamefont {Li},
  \citenamefont {Zaki}, \citenamefont {Garlea}, \citenamefont {Savici},
  \citenamefont {Fobes}, \citenamefont {Xu}, \citenamefont {Camino},
  \citenamefont {Petrovic}, \citenamefont {Gu}, \citenamefont {Johnson} \emph
  {et~al.}}]{li2021electronic}%
  \BibitemOpen
  \bibfield  {author} {\bibinfo {author} {\bibfnamefont {Y.}~\bibnamefont
  {Li}}, \bibinfo {author} {\bibfnamefont {N.}~\bibnamefont {Zaki}}, \bibinfo
  {author} {\bibfnamefont {V.~O.}\ \bibnamefont {Garlea}}, \bibinfo {author}
  {\bibfnamefont {A.~T.}\ \bibnamefont {Savici}}, \bibinfo {author}
  {\bibfnamefont {D.}~\bibnamefont {Fobes}}, \bibinfo {author} {\bibfnamefont
  {Z.}~\bibnamefont {Xu}}, \bibinfo {author} {\bibfnamefont {F.}~\bibnamefont
  {Camino}}, \bibinfo {author} {\bibfnamefont {C.}~\bibnamefont {Petrovic}},
  \bibinfo {author} {\bibfnamefont {G.}~\bibnamefont {Gu}}, \bibinfo {author}
  {\bibfnamefont {P.~D.}\ \bibnamefont {Johnson}},  \emph {et~al.},\
  }\href@noop {} {\bibfield  {journal} {\bibinfo  {journal} {Nature Materials}\
  }\textbf {\bibinfo {volume} {20}},\ \bibinfo {pages} {1221} (\bibinfo {year}
  {2021})}\BibitemShut {NoStop}%
\bibitem [{\citenamefont {Anderson}(1973)}]{anderson1973resonating}%
  \BibitemOpen
  \bibfield  {author} {\bibinfo {author} {\bibfnamefont {P.~W.}\ \bibnamefont
  {Anderson}},\ }\href@noop {} {\bibfield  {journal} {\bibinfo  {journal}
  {Materials Research Bulletin}\ }\textbf {\bibinfo {volume} {8}},\ \bibinfo
  {pages} {153} (\bibinfo {year} {1973})}\BibitemShut {NoStop}%
\bibitem [{\citenamefont {Kitaev}(2006)}]{kitaev2006anyons}%
  \BibitemOpen
  \bibfield  {author} {\bibinfo {author} {\bibfnamefont {A.}~\bibnamefont
  {Kitaev}},\ }\href@noop {} {\bibfield  {journal} {\bibinfo  {journal} {Annals
  of Physics}\ }\textbf {\bibinfo {volume} {321}},\ \bibinfo {pages} {2}
  (\bibinfo {year} {2006})}\BibitemShut {NoStop}%
\bibitem [{\citenamefont {Huang}\ and\ \citenamefont
  {Sheng}(2022)}]{huang2022topological}%
  \BibitemOpen
  \bibfield  {author} {\bibinfo {author} {\bibfnamefont {Y.}~\bibnamefont
  {Huang}}\ and\ \bibinfo {author} {\bibfnamefont {D.}~\bibnamefont {Sheng}},\
  }\href@noop {} {\bibfield  {journal} {\bibinfo  {journal} {Physical Review
  X}\ }\textbf {\bibinfo {volume} {12}},\ \bibinfo {pages} {031009} (\bibinfo
  {year} {2022})}\BibitemShut {NoStop}%
\bibitem [{\citenamefont {Balents}(2010)}]{balents2010spin}%
  \BibitemOpen
  \bibfield  {author} {\bibinfo {author} {\bibfnamefont {L.}~\bibnamefont
  {Balents}},\ }\href@noop {} {\bibfield  {journal} {\bibinfo  {journal}
  {Nature}\ }\textbf {\bibinfo {volume} {464}},\ \bibinfo {pages} {199}
  (\bibinfo {year} {2010})}\BibitemShut {NoStop}%
\bibitem [{\citenamefont {Savary}\ and\ \citenamefont
  {Balents}(2016)}]{savary2016quantum}%
  \BibitemOpen
  \bibfield  {author} {\bibinfo {author} {\bibfnamefont {L.}~\bibnamefont
  {Savary}}\ and\ \bibinfo {author} {\bibfnamefont {L.}~\bibnamefont
  {Balents}},\ }\href@noop {} {\bibfield  {journal} {\bibinfo  {journal}
  {Reports on Progress in Physics}\ }\textbf {\bibinfo {volume} {80}},\
  \bibinfo {pages} {016502} (\bibinfo {year} {2016})}\BibitemShut {NoStop}%
\bibitem [{\citenamefont {Knolle}\ and\ \citenamefont
  {Moessner}(2019)}]{knolle2019field}%
  \BibitemOpen
  \bibfield  {author} {\bibinfo {author} {\bibfnamefont {J.}~\bibnamefont
  {Knolle}}\ and\ \bibinfo {author} {\bibfnamefont {R.}~\bibnamefont
  {Moessner}},\ }\href@noop {} {\bibfield  {journal} {\bibinfo  {journal}
  {Annual Review of Condensed Matter Physics}\ }\textbf {\bibinfo {volume}
  {10}},\ \bibinfo {pages} {451} (\bibinfo {year} {2019})}\BibitemShut
  {NoStop}%
\bibitem [{\citenamefont {Xu}\ \emph {et~al.}(2020)\citenamefont {Xu},
  \citenamefont {Feng}, \citenamefont {Kawamura}, \citenamefont {Yamaji},
  \citenamefont {Nahas}, \citenamefont {Prokhorenko}, \citenamefont {Qi},
  \citenamefont {Xiang},\ and\ \citenamefont {Bellaiche}}]{xu2020possible}%
  \BibitemOpen
  \bibfield  {author} {\bibinfo {author} {\bibfnamefont {C.}~\bibnamefont
  {Xu}}, \bibinfo {author} {\bibfnamefont {J.}~\bibnamefont {Feng}}, \bibinfo
  {author} {\bibfnamefont {M.}~\bibnamefont {Kawamura}}, \bibinfo {author}
  {\bibfnamefont {Y.}~\bibnamefont {Yamaji}}, \bibinfo {author} {\bibfnamefont
  {Y.}~\bibnamefont {Nahas}}, \bibinfo {author} {\bibfnamefont
  {S.}~\bibnamefont {Prokhorenko}}, \bibinfo {author} {\bibfnamefont
  {Y.}~\bibnamefont {Qi}}, \bibinfo {author} {\bibfnamefont {H.}~\bibnamefont
  {Xiang}}, \ and\ \bibinfo {author} {\bibfnamefont {L.}~\bibnamefont
  {Bellaiche}},\ }\href@noop {} {\bibfield  {journal} {\bibinfo  {journal}
  {Physical Review Letters}\ }\textbf {\bibinfo {volume} {124}},\ \bibinfo
  {pages} {087205} (\bibinfo {year} {2020})}\BibitemShut {NoStop}%
\bibitem [{\citenamefont {Ji}\ \emph {et~al.}(2021)\citenamefont {Ji},
  \citenamefont {Sun}, \citenamefont {Cai}, \citenamefont {Wang}, \citenamefont
  {Sun}, \citenamefont {Ren}, \citenamefont {Zhang}, \citenamefont {Jin},\ and\
  \citenamefont {Zhang}}]{ji2021rare}%
  \BibitemOpen
  \bibfield  {author} {\bibinfo {author} {\bibfnamefont {J.}~\bibnamefont
  {Ji}}, \bibinfo {author} {\bibfnamefont {M.}~\bibnamefont {Sun}}, \bibinfo
  {author} {\bibfnamefont {Y.}~\bibnamefont {Cai}}, \bibinfo {author}
  {\bibfnamefont {Y.}~\bibnamefont {Wang}}, \bibinfo {author} {\bibfnamefont
  {Y.}~\bibnamefont {Sun}}, \bibinfo {author} {\bibfnamefont {W.}~\bibnamefont
  {Ren}}, \bibinfo {author} {\bibfnamefont {Z.}~\bibnamefont {Zhang}}, \bibinfo
  {author} {\bibfnamefont {F.}~\bibnamefont {Jin}}, \ and\ \bibinfo {author}
  {\bibfnamefont {Q.}~\bibnamefont {Zhang}},\ }\href@noop {} {\bibfield
  {journal} {\bibinfo  {journal} {Chinese Physics Letters}\ }\textbf {\bibinfo
  {volume} {38}},\ \bibinfo {pages} {047502} (\bibinfo {year}
  {2021})}\BibitemShut {NoStop}%
\bibitem [{\citenamefont {Loidl}\ \emph {et~al.}(2021)\citenamefont {Loidl},
  \citenamefont {Lunkenheimer},\ and\ \citenamefont
  {Tsurkan}}]{loidl2021proximate}%
  \BibitemOpen
  \bibfield  {author} {\bibinfo {author} {\bibfnamefont {A.}~\bibnamefont
  {Loidl}}, \bibinfo {author} {\bibfnamefont {P.}~\bibnamefont {Lunkenheimer}},
  \ and\ \bibinfo {author} {\bibfnamefont {V.}~\bibnamefont {Tsurkan}},\
  }\href@noop {} {\bibfield  {journal} {\bibinfo  {journal} {Journal of
  Physics: Condensed Matter}\ } (\bibinfo {year} {2021})}\BibitemShut {NoStop}%
\bibitem [{\citenamefont {Takagi}\ \emph {et~al.}(2019)\citenamefont {Takagi},
  \citenamefont {Takayama}, \citenamefont {Jackeli}, \citenamefont
  {Khaliullin},\ and\ \citenamefont {Nagler}}]{takagi2019concept}%
  \BibitemOpen
  \bibfield  {author} {\bibinfo {author} {\bibfnamefont {H.}~\bibnamefont
  {Takagi}}, \bibinfo {author} {\bibfnamefont {T.}~\bibnamefont {Takayama}},
  \bibinfo {author} {\bibfnamefont {G.}~\bibnamefont {Jackeli}}, \bibinfo
  {author} {\bibfnamefont {G.}~\bibnamefont {Khaliullin}}, \ and\ \bibinfo
  {author} {\bibfnamefont {S.~E.}\ \bibnamefont {Nagler}},\ }\href@noop {}
  {\bibfield  {journal} {\bibinfo  {journal} {Nature Reviews Physics}\ }\textbf
  {\bibinfo {volume} {1}},\ \bibinfo {pages} {264} (\bibinfo {year}
  {2019})}\BibitemShut {NoStop}%
\bibitem [{\citenamefont {Burch}\ \emph {et~al.}(2018)\citenamefont {Burch},
  \citenamefont {Mandrus},\ and\ \citenamefont {Park}}]{burch2018magnetism}%
  \BibitemOpen
  \bibfield  {author} {\bibinfo {author} {\bibfnamefont {K.~S.}\ \bibnamefont
  {Burch}}, \bibinfo {author} {\bibfnamefont {D.}~\bibnamefont {Mandrus}}, \
  and\ \bibinfo {author} {\bibfnamefont {J.-G.}\ \bibnamefont {Park}},\
  }\href@noop {} {\bibfield  {journal} {\bibinfo  {journal} {Nature}\ }\textbf
  {\bibinfo {volume} {563}},\ \bibinfo {pages} {47} (\bibinfo {year}
  {2018})}\BibitemShut {NoStop}%
\bibitem [{\citenamefont {Fei}\ \emph {et~al.}(2018)\citenamefont {Fei},
  \citenamefont {Huang}, \citenamefont {Malinowski}, \citenamefont {Wang},
  \citenamefont {Song}, \citenamefont {Sanchez}, \citenamefont {Yao},
  \citenamefont {Xiao}, \citenamefont {Zhu}, \citenamefont {May} \emph
  {et~al.}}]{fei2018two}%
  \BibitemOpen
  \bibfield  {author} {\bibinfo {author} {\bibfnamefont {Z.}~\bibnamefont
  {Fei}}, \bibinfo {author} {\bibfnamefont {B.}~\bibnamefont {Huang}}, \bibinfo
  {author} {\bibfnamefont {P.}~\bibnamefont {Malinowski}}, \bibinfo {author}
  {\bibfnamefont {W.}~\bibnamefont {Wang}}, \bibinfo {author} {\bibfnamefont
  {T.}~\bibnamefont {Song}}, \bibinfo {author} {\bibfnamefont {J.}~\bibnamefont
  {Sanchez}}, \bibinfo {author} {\bibfnamefont {W.}~\bibnamefont {Yao}},
  \bibinfo {author} {\bibfnamefont {D.}~\bibnamefont {Xiao}}, \bibinfo {author}
  {\bibfnamefont {X.}~\bibnamefont {Zhu}}, \bibinfo {author} {\bibfnamefont
  {A.~F.}\ \bibnamefont {May}},  \emph {et~al.},\ }\href@noop {} {\bibfield
  {journal} {\bibinfo  {journal} {Nature Materials}\ }\textbf {\bibinfo
  {volume} {17}},\ \bibinfo {pages} {778} (\bibinfo {year} {2018})}\BibitemShut
  {NoStop}%
\bibitem [{\citenamefont {Xu}\ \emph {et~al.}(2022)\citenamefont {Xu},
  \citenamefont {Ray}, \citenamefont {Shao}, \citenamefont {Jiang},
  \citenamefont {Lee}, \citenamefont {Weber}, \citenamefont {Goldberger},
  \citenamefont {Watanabe}, \citenamefont {Taniguchi}, \citenamefont {Muller}
  \emph {et~al.}}]{xu2022coexisting}%
  \BibitemOpen
  \bibfield  {author} {\bibinfo {author} {\bibfnamefont {Y.}~\bibnamefont
  {Xu}}, \bibinfo {author} {\bibfnamefont {A.}~\bibnamefont {Ray}}, \bibinfo
  {author} {\bibfnamefont {Y.-T.}\ \bibnamefont {Shao}}, \bibinfo {author}
  {\bibfnamefont {S.}~\bibnamefont {Jiang}}, \bibinfo {author} {\bibfnamefont
  {K.}~\bibnamefont {Lee}}, \bibinfo {author} {\bibfnamefont {D.}~\bibnamefont
  {Weber}}, \bibinfo {author} {\bibfnamefont {J.~E.}\ \bibnamefont
  {Goldberger}}, \bibinfo {author} {\bibfnamefont {K.}~\bibnamefont
  {Watanabe}}, \bibinfo {author} {\bibfnamefont {T.}~\bibnamefont {Taniguchi}},
  \bibinfo {author} {\bibfnamefont {D.~A.}\ \bibnamefont {Muller}},  \emph
  {et~al.},\ }\href@noop {} {\bibfield  {journal} {\bibinfo  {journal} {Nature
  Nanotechnology}\ }\textbf {\bibinfo {volume} {17}},\ \bibinfo {pages} {143}
  (\bibinfo {year} {2022})}\BibitemShut {NoStop}%
\bibitem [{\citenamefont {Jiang}\ \emph {et~al.}(2018)\citenamefont {Jiang},
  \citenamefont {Li}, \citenamefont {Wang}, \citenamefont {Mak},\ and\
  \citenamefont {Shan}}]{jiang2018controlling}%
  \BibitemOpen
  \bibfield  {author} {\bibinfo {author} {\bibfnamefont {S.}~\bibnamefont
  {Jiang}}, \bibinfo {author} {\bibfnamefont {L.}~\bibnamefont {Li}}, \bibinfo
  {author} {\bibfnamefont {Z.}~\bibnamefont {Wang}}, \bibinfo {author}
  {\bibfnamefont {K.~F.}\ \bibnamefont {Mak}}, \ and\ \bibinfo {author}
  {\bibfnamefont {J.}~\bibnamefont {Shan}},\ }\href@noop {} {\bibfield
  {journal} {\bibinfo  {journal} {Nature Nanotechnology}\ }\textbf {\bibinfo
  {volume} {13}},\ \bibinfo {pages} {549} (\bibinfo {year} {2018})}\BibitemShut
  {NoStop}%
\bibitem [{\citenamefont {Afanasiev}\ \emph {et~al.}(2021)\citenamefont
  {Afanasiev}, \citenamefont {Hortensius}, \citenamefont {Matthiesen},
  \citenamefont {Ma{\~n}as-Valero}, \citenamefont {{\v{S}}i{\v{s}}kins},
  \citenamefont {Lee}, \citenamefont {Lesne}, \citenamefont {van Der~Zant},
  \citenamefont {Steeneken}, \citenamefont {Ivanov} \emph
  {et~al.}}]{afanasiev2021controlling}%
  \BibitemOpen
  \bibfield  {author} {\bibinfo {author} {\bibfnamefont {D.}~\bibnamefont
  {Afanasiev}}, \bibinfo {author} {\bibfnamefont {J.~R.}\ \bibnamefont
  {Hortensius}}, \bibinfo {author} {\bibfnamefont {M.}~\bibnamefont
  {Matthiesen}}, \bibinfo {author} {\bibfnamefont {S.}~\bibnamefont
  {Ma{\~n}as-Valero}}, \bibinfo {author} {\bibfnamefont {M.}~\bibnamefont
  {{\v{S}}i{\v{s}}kins}}, \bibinfo {author} {\bibfnamefont {M.}~\bibnamefont
  {Lee}}, \bibinfo {author} {\bibfnamefont {E.}~\bibnamefont {Lesne}}, \bibinfo
  {author} {\bibfnamefont {H.~S.}\ \bibnamefont {van Der~Zant}}, \bibinfo
  {author} {\bibfnamefont {P.~G.}\ \bibnamefont {Steeneken}}, \bibinfo {author}
  {\bibfnamefont {B.~A.}\ \bibnamefont {Ivanov}},  \emph {et~al.},\ }\href@noop
  {} {\bibfield  {journal} {\bibinfo  {journal} {Science Advances}\ }\textbf
  {\bibinfo {volume} {7}},\ \bibinfo {pages} {eabf3096} (\bibinfo {year}
  {2021})}\BibitemShut {NoStop}%
\bibitem [{\citenamefont {Wang}\ and\ \citenamefont
  {Astruc}(2019)}]{wang2019state}%
  \BibitemOpen
  \bibfield  {author} {\bibinfo {author} {\bibfnamefont {Q.}~\bibnamefont
  {Wang}}\ and\ \bibinfo {author} {\bibfnamefont {D.}~\bibnamefont {Astruc}},\
  }\href@noop {} {\bibfield  {journal} {\bibinfo  {journal} {Chemical Reviews}\
  }\textbf {\bibinfo {volume} {120}},\ \bibinfo {pages} {1438} (\bibinfo {year}
  {2019})}\BibitemShut {NoStop}%
\bibitem [{\citenamefont {Yamada}\ \emph {et~al.}(2017)\citenamefont {Yamada},
  \citenamefont {Fujita},\ and\ \citenamefont
  {Oshikawa}}]{yamada2017designing}%
  \BibitemOpen
  \bibfield  {author} {\bibinfo {author} {\bibfnamefont {M.~G.}\ \bibnamefont
  {Yamada}}, \bibinfo {author} {\bibfnamefont {H.}~\bibnamefont {Fujita}}, \
  and\ \bibinfo {author} {\bibfnamefont {M.}~\bibnamefont {Oshikawa}},\
  }\href@noop {} {\bibfield  {journal} {\bibinfo  {journal} {Physical Review
  Letters}\ }\textbf {\bibinfo {volume} {119}},\ \bibinfo {pages} {057202}
  (\bibinfo {year} {2017})}\BibitemShut {NoStop}%
\bibitem [{\citenamefont {Murphy}\ \emph {et~al.}(2021)\citenamefont {Murphy},
  \citenamefont {Darago}, \citenamefont {Ziebel}, \citenamefont {Peterson},
  \citenamefont {Zaia}, \citenamefont {Mara}, \citenamefont {Lussier},
  \citenamefont {Velasquez}, \citenamefont {Shuh}, \citenamefont {Urban} \emph
  {et~al.}}]{murphy2021exchange}%
  \BibitemOpen
  \bibfield  {author} {\bibinfo {author} {\bibfnamefont {R.~A.}\ \bibnamefont
  {Murphy}}, \bibinfo {author} {\bibfnamefont {L.~E.}\ \bibnamefont {Darago}},
  \bibinfo {author} {\bibfnamefont {M.~E.}\ \bibnamefont {Ziebel}}, \bibinfo
  {author} {\bibfnamefont {E.~A.}\ \bibnamefont {Peterson}}, \bibinfo {author}
  {\bibfnamefont {E.~W.}\ \bibnamefont {Zaia}}, \bibinfo {author}
  {\bibfnamefont {M.~W.}\ \bibnamefont {Mara}}, \bibinfo {author}
  {\bibfnamefont {D.}~\bibnamefont {Lussier}}, \bibinfo {author} {\bibfnamefont
  {E.~O.}\ \bibnamefont {Velasquez}}, \bibinfo {author} {\bibfnamefont {D.~K.}\
  \bibnamefont {Shuh}}, \bibinfo {author} {\bibfnamefont {J.~J.}\ \bibnamefont
  {Urban}},  \emph {et~al.},\ }\href@noop {} {\bibfield  {journal} {\bibinfo
  {journal} {ACS central science}\ }\textbf {\bibinfo {volume} {7}},\ \bibinfo
  {pages} {1317} (\bibinfo {year} {2021})}\BibitemShut {NoStop}%
\bibitem [{\citenamefont {Jiang}\ \emph {et~al.}(2021)\citenamefont {Jiang},
  \citenamefont {Ni},\ and\ \citenamefont {Liu}}]{jiang2021exotic}%
  \BibitemOpen
  \bibfield  {author} {\bibinfo {author} {\bibfnamefont {W.}~\bibnamefont
  {Jiang}}, \bibinfo {author} {\bibfnamefont {X.}~\bibnamefont {Ni}}, \ and\
  \bibinfo {author} {\bibfnamefont {F.}~\bibnamefont {Liu}},\ }\href@noop {}
  {\bibfield  {journal} {\bibinfo  {journal} {Accounts of Chemical Research}\
  }\textbf {\bibinfo {volume} {54}},\ \bibinfo {pages} {416} (\bibinfo {year}
  {2021})}\BibitemShut {NoStop}%
\bibitem [{\citenamefont {Kresse}\ and\ \citenamefont
  {Joubert}(1999)}]{Kresse1999}%
  \BibitemOpen
  \bibfield  {author} {\bibinfo {author} {\bibfnamefont {G.}~\bibnamefont
  {Kresse}}\ and\ \bibinfo {author} {\bibfnamefont {D.}~\bibnamefont
  {Joubert}},\ }\href@noop {} {\bibfield  {journal} {\bibinfo  {journal} {Phys.
  Rev. B}\ }\textbf {\bibinfo {volume} {59}},\ \bibinfo {pages} {1758}
  (\bibinfo {year} {1999})}\BibitemShut {NoStop}%
\bibitem [{\citenamefont {Kresse}\ and\ \citenamefont
  {Furthm{\"{u}}ller}(1996)}]{Kresse1996}%
  \BibitemOpen
  \bibfield  {author} {\bibinfo {author} {\bibfnamefont {G.}~\bibnamefont
  {Kresse}}\ and\ \bibinfo {author} {\bibfnamefont {J.}~\bibnamefont
  {Furthm{\"{u}}ller}},\ }\href@noop {} {\bibfield  {journal} {\bibinfo
  {journal} {Phys. Rev. B}\ }\textbf {\bibinfo {volume} {54}},\ \bibinfo
  {pages} {11169} (\bibinfo {year} {1996})}\BibitemShut {NoStop}%
\bibitem [{\citenamefont {Kresse}\ and\ \citenamefont
  {Hafner}(1993)}]{Kresse1993}%
  \BibitemOpen
  \bibfield  {author} {\bibinfo {author} {\bibfnamefont {G.}~\bibnamefont
  {Kresse}}\ and\ \bibinfo {author} {\bibfnamefont {J.}~\bibnamefont
  {Hafner}},\ }\href@noop {} {\bibfield  {journal} {\bibinfo  {journal} {Phys.
  Rev. B}\ }\textbf {\bibinfo {volume} {48}},\ \bibinfo {pages} {13115}
  (\bibinfo {year} {1993})}\BibitemShut {NoStop}%
\bibitem [{\citenamefont {Perdew}\ \emph {et~al.}(1996)\citenamefont {Perdew},
  \citenamefont {Burke},\ and\ \citenamefont
  {Ernzerhof}}]{perdew1996generalized}%
  \BibitemOpen
  \bibfield  {author} {\bibinfo {author} {\bibfnamefont {J.~P.}\ \bibnamefont
  {Perdew}}, \bibinfo {author} {\bibfnamefont {K.}~\bibnamefont {Burke}}, \
  and\ \bibinfo {author} {\bibfnamefont {M.}~\bibnamefont {Ernzerhof}},\
  }\href@noop {} {\bibfield  {journal} {\bibinfo  {journal} {Physical Review
  Letters}\ }\textbf {\bibinfo {volume} {77}},\ \bibinfo {pages} {3865}
  (\bibinfo {year} {1996})}\BibitemShut {NoStop}%
\bibitem [{\citenamefont {Jeon}\ \emph {et~al.}(2015)\citenamefont {Jeon},
  \citenamefont {Negru}, \citenamefont {Van~Duyne},\ and\ \citenamefont
  {Harris}}]{jeon20152d}%
  \BibitemOpen
  \bibfield  {author} {\bibinfo {author} {\bibfnamefont {I.-R.}\ \bibnamefont
  {Jeon}}, \bibinfo {author} {\bibfnamefont {B.}~\bibnamefont {Negru}},
  \bibinfo {author} {\bibfnamefont {R.~P.}\ \bibnamefont {Van~Duyne}}, \ and\
  \bibinfo {author} {\bibfnamefont {T.~D.}\ \bibnamefont {Harris}},\
  }\href@noop {} {\bibfield  {journal} {\bibinfo  {journal} {Journal of the
  American Chemical Society}\ }\textbf {\bibinfo {volume} {137}},\ \bibinfo
  {pages} {15699} (\bibinfo {year} {2015})}\BibitemShut {NoStop}%
\bibitem [{\citenamefont {White}(1992)}]{white1992density}%
  \BibitemOpen
  \bibfield  {author} {\bibinfo {author} {\bibfnamefont {S.~R.}\ \bibnamefont
  {White}},\ }\href@noop {} {\bibfield  {journal} {\bibinfo  {journal}
  {Physical Review Letters}\ }\textbf {\bibinfo {volume} {69}},\ \bibinfo
  {pages} {2863} (\bibinfo {year} {1992})}\BibitemShut {NoStop}%
\bibitem [{\citenamefont {White}(1993)}]{white1993density}%
  \BibitemOpen
  \bibfield  {author} {\bibinfo {author} {\bibfnamefont {S.~R.}\ \bibnamefont
  {White}},\ }\href@noop {} {\bibfield  {journal} {\bibinfo  {journal}
  {Physical Review B}\ }\textbf {\bibinfo {volume} {48}},\ \bibinfo {pages}
  {10345} (\bibinfo {year} {1993})}\BibitemShut {NoStop}%
\bibitem [{\citenamefont {Schollw{\"o}ck}(2011)}]{schollwock2011density}%
  \BibitemOpen
  \bibfield  {author} {\bibinfo {author} {\bibfnamefont {U.}~\bibnamefont
  {Schollw{\"o}ck}},\ }\href@noop {} {\bibfield  {journal} {\bibinfo  {journal}
  {Annals of physics}\ }\textbf {\bibinfo {volume} {326}},\ \bibinfo {pages}
  {96} (\bibinfo {year} {2011})}\BibitemShut {NoStop}%
\bibitem [{\citenamefont {Hauschild}\ and\ \citenamefont
  {Pollmann}(2018)}]{hauschild2018efficient}%
  \BibitemOpen
  \bibfield  {author} {\bibinfo {author} {\bibfnamefont {J.}~\bibnamefont
  {Hauschild}}\ and\ \bibinfo {author} {\bibfnamefont {F.}~\bibnamefont
  {Pollmann}},\ }\href@noop {} {\bibfield  {journal} {\bibinfo  {journal}
  {SciPost Physics Lecture Notes}\ ,\ \bibinfo {pages} {005}} (\bibinfo {year}
  {2018})}\BibitemShut {NoStop}%
\bibitem [{\citenamefont {Xiang}\ \emph {et~al.}(2013)\citenamefont {Xiang},
  \citenamefont {Lee}, \citenamefont {Koo}, \citenamefont {Gong},\ and\
  \citenamefont {Whangbo}}]{xiang2013magnetic}%
  \BibitemOpen
  \bibfield  {author} {\bibinfo {author} {\bibfnamefont {H.}~\bibnamefont
  {Xiang}}, \bibinfo {author} {\bibfnamefont {C.}~\bibnamefont {Lee}}, \bibinfo
  {author} {\bibfnamefont {H.-J.}\ \bibnamefont {Koo}}, \bibinfo {author}
  {\bibfnamefont {X.}~\bibnamefont {Gong}}, \ and\ \bibinfo {author}
  {\bibfnamefont {M.-H.}\ \bibnamefont {Whangbo}},\ }\href@noop {} {\bibfield
  {journal} {\bibinfo  {journal} {Dalton Transactions}\ }\textbf {\bibinfo
  {volume} {42}},\ \bibinfo {pages} {823} (\bibinfo {year} {2013})}\BibitemShut
  {NoStop}%
\bibitem [{Sup()}]{SuppMaterial}%
  \BibitemOpen
  \href@noop {} {}\bibinfo {howpublished} {See Supplemental Materials at [URL
  will be inserted by publisher] for more details.}\BibitemShut {Stop}%
\bibitem [{\citenamefont {Watanabe}\ \emph {et~al.}(2022)\citenamefont
  {Watanabe}, \citenamefont {Trebst},\ and\ \citenamefont
  {Hickey}}]{watanabe2022frustrated}%
  \BibitemOpen
  \bibfield  {author} {\bibinfo {author} {\bibfnamefont {Y.}~\bibnamefont
  {Watanabe}}, \bibinfo {author} {\bibfnamefont {S.}~\bibnamefont {Trebst}}, \
  and\ \bibinfo {author} {\bibfnamefont {C.}~\bibnamefont {Hickey}},\
  }\href@noop {} {\bibfield  {journal} {\bibinfo  {journal} {arXiv preprint
  arXiv:2212.14053}\ } (\bibinfo {year} {2022})}\BibitemShut {NoStop}%
\bibitem [{\citenamefont {Bose}\ \emph {et~al.}(2022)\citenamefont {Bose},
  \citenamefont {Routh}, \citenamefont {Voleti}, \citenamefont {Saha},
  \citenamefont {Kumar}, \citenamefont {Saha-Dasgupta},\ and\ \citenamefont
  {Paramekanti}}]{bose2022proximate}%
  \BibitemOpen
  \bibfield  {author} {\bibinfo {author} {\bibfnamefont {A.}~\bibnamefont
  {Bose}}, \bibinfo {author} {\bibfnamefont {M.}~\bibnamefont {Routh}},
  \bibinfo {author} {\bibfnamefont {S.}~\bibnamefont {Voleti}}, \bibinfo
  {author} {\bibfnamefont {S.~K.}\ \bibnamefont {Saha}}, \bibinfo {author}
  {\bibfnamefont {M.}~\bibnamefont {Kumar}}, \bibinfo {author} {\bibfnamefont
  {T.}~\bibnamefont {Saha-Dasgupta}}, \ and\ \bibinfo {author} {\bibfnamefont
  {A.}~\bibnamefont {Paramekanti}},\ }\href@noop {} {\bibfield  {journal}
  {\bibinfo  {journal} {arXiv preprint arXiv:2212.13271}\ } (\bibinfo {year}
  {2022})}\BibitemShut {NoStop}%
\bibitem [{\citenamefont {Jiang}\ \emph {et~al.}(2023)\citenamefont {Jiang},
  \citenamefont {White},\ and\ \citenamefont {Chernyshev}}]{jiang2023quantum}%
  \BibitemOpen
  \bibfield  {author} {\bibinfo {author} {\bibfnamefont {S.}~\bibnamefont
  {Jiang}}, \bibinfo {author} {\bibfnamefont {S.~R.}\ \bibnamefont {White}}, \
  and\ \bibinfo {author} {\bibfnamefont {A.}~\bibnamefont {Chernyshev}},\
  }\href@noop {} {\bibfield  {journal} {\bibinfo  {journal} {arXiv preprint
  arXiv:2304.06062}\ } (\bibinfo {year} {2023})}\BibitemShut {NoStop}%
\bibitem [{\citenamefont {Wildes}\ \emph {et~al.}(1994)\citenamefont {Wildes},
  \citenamefont {Kennedy},\ and\ \citenamefont {Hicks}}]{wildes1994true}%
  \BibitemOpen
  \bibfield  {author} {\bibinfo {author} {\bibfnamefont {A.}~\bibnamefont
  {Wildes}}, \bibinfo {author} {\bibfnamefont {S.}~\bibnamefont {Kennedy}}, \
  and\ \bibinfo {author} {\bibfnamefont {T.}~\bibnamefont {Hicks}},\
  }\href@noop {} {\bibfield  {journal} {\bibinfo  {journal} {Journal of
  Physics: Condensed Matter}\ }\textbf {\bibinfo {volume} {6}},\ \bibinfo
  {pages} {L335} (\bibinfo {year} {1994})}\BibitemShut {NoStop}%
\bibitem [{\citenamefont {Ni}\ \emph {et~al.}(2020)\citenamefont {Ni},
  \citenamefont {Zhou}, \citenamefont {Sethi},\ and\ \citenamefont
  {Liu}}]{ni2020pi}%
  \BibitemOpen
  \bibfield  {author} {\bibinfo {author} {\bibfnamefont {X.}~\bibnamefont
  {Ni}}, \bibinfo {author} {\bibfnamefont {Y.}~\bibnamefont {Zhou}}, \bibinfo
  {author} {\bibfnamefont {G.}~\bibnamefont {Sethi}}, \ and\ \bibinfo {author}
  {\bibfnamefont {F.}~\bibnamefont {Liu}},\ }\href@noop {} {\bibfield
  {journal} {\bibinfo  {journal} {Physical Chemistry Chemical Physics}\
  }\textbf {\bibinfo {volume} {22}},\ \bibinfo {pages} {25827} (\bibinfo {year}
  {2020})}\BibitemShut {NoStop}%
\bibitem [{\citenamefont {Sethi}\ \emph {et~al.}(2021)\citenamefont {Sethi},
  \citenamefont {Zhou}, \citenamefont {Zhu}, \citenamefont {Yang},\ and\
  \citenamefont {Liu}}]{sethi2021flat}%
  \BibitemOpen
  \bibfield  {author} {\bibinfo {author} {\bibfnamefont {G.}~\bibnamefont
  {Sethi}}, \bibinfo {author} {\bibfnamefont {Y.}~\bibnamefont {Zhou}},
  \bibinfo {author} {\bibfnamefont {L.}~\bibnamefont {Zhu}}, \bibinfo {author}
  {\bibfnamefont {L.}~\bibnamefont {Yang}}, \ and\ \bibinfo {author}
  {\bibfnamefont {F.}~\bibnamefont {Liu}},\ }\href@noop {} {\bibfield
  {journal} {\bibinfo  {journal} {Physical Review Letters}\ }\textbf {\bibinfo
  {volume} {126}},\ \bibinfo {pages} {196403} (\bibinfo {year}
  {2021})}\BibitemShut {NoStop}%
\bibitem [{\citenamefont {Zhou}\ \emph {et~al.}(2019)\citenamefont {Zhou},
  \citenamefont {Sethi}, \citenamefont {Liu}, \citenamefont {Wang},\ and\
  \citenamefont {Liu}}]{zhou2019excited}%
  \BibitemOpen
  \bibfield  {author} {\bibinfo {author} {\bibfnamefont {Y.}~\bibnamefont
  {Zhou}}, \bibinfo {author} {\bibfnamefont {G.}~\bibnamefont {Sethi}},
  \bibinfo {author} {\bibfnamefont {H.}~\bibnamefont {Liu}}, \bibinfo {author}
  {\bibfnamefont {Z.}~\bibnamefont {Wang}}, \ and\ \bibinfo {author}
  {\bibfnamefont {F.}~\bibnamefont {Liu}},\ }\href@noop {} {\bibfield
  {journal} {\bibinfo  {journal} {arXiv preprint arXiv:1908.03689}\ } (\bibinfo
  {year} {2019})}\BibitemShut {NoStop}%
\bibitem [{\citenamefont {Cao}\ \emph {et~al.}(2018)\citenamefont {Cao},
  \citenamefont {Fatemi}, \citenamefont {Demir}, \citenamefont {Fang},
  \citenamefont {Tomarken}, \citenamefont {Luo}, \citenamefont
  {Sanchez-Yamagishi}, \citenamefont {Watanabe}, \citenamefont {Taniguchi},
  \citenamefont {Kaxiras} \emph {et~al.}}]{cao2018correlated}%
  \BibitemOpen
  \bibfield  {author} {\bibinfo {author} {\bibfnamefont {Y.}~\bibnamefont
  {Cao}}, \bibinfo {author} {\bibfnamefont {V.}~\bibnamefont {Fatemi}},
  \bibinfo {author} {\bibfnamefont {A.}~\bibnamefont {Demir}}, \bibinfo
  {author} {\bibfnamefont {S.}~\bibnamefont {Fang}}, \bibinfo {author}
  {\bibfnamefont {S.~L.}\ \bibnamefont {Tomarken}}, \bibinfo {author}
  {\bibfnamefont {J.~Y.}\ \bibnamefont {Luo}}, \bibinfo {author} {\bibfnamefont
  {J.~D.}\ \bibnamefont {Sanchez-Yamagishi}}, \bibinfo {author} {\bibfnamefont
  {K.}~\bibnamefont {Watanabe}}, \bibinfo {author} {\bibfnamefont
  {T.}~\bibnamefont {Taniguchi}}, \bibinfo {author} {\bibfnamefont
  {E.}~\bibnamefont {Kaxiras}},  \emph {et~al.},\ }\href@noop {} {\bibfield
  {journal} {\bibinfo  {journal} {Nature}\ }\textbf {\bibinfo {volume} {556}},\
  \bibinfo {pages} {80} (\bibinfo {year} {2018})}\BibitemShut {NoStop}%
\bibitem [{\citenamefont {Pertsova}\ and\ \citenamefont
  {Balatsky}(2020)}]{pertsova2020dynamically}%
  \BibitemOpen
  \bibfield  {author} {\bibinfo {author} {\bibfnamefont {A.}~\bibnamefont
  {Pertsova}}\ and\ \bibinfo {author} {\bibfnamefont {A.~V.}\ \bibnamefont
  {Balatsky}},\ }\href@noop {} {\bibfield  {journal} {\bibinfo  {journal}
  {Annalen der Physik}\ }\textbf {\bibinfo {volume} {532}},\ \bibinfo {pages}
  {1900549} (\bibinfo {year} {2020})}\BibitemShut {NoStop}%
\bibitem [{\citenamefont {Kotov}\ \emph {et~al.}(2012)\citenamefont {Kotov},
  \citenamefont {Uchoa}, \citenamefont {Pereira}, \citenamefont {Guinea},\ and\
  \citenamefont {Neto}}]{kotov2012electron}%
  \BibitemOpen
  \bibfield  {author} {\bibinfo {author} {\bibfnamefont {V.~N.}\ \bibnamefont
  {Kotov}}, \bibinfo {author} {\bibfnamefont {B.}~\bibnamefont {Uchoa}},
  \bibinfo {author} {\bibfnamefont {V.~M.}\ \bibnamefont {Pereira}}, \bibinfo
  {author} {\bibfnamefont {F.}~\bibnamefont {Guinea}}, \ and\ \bibinfo {author}
  {\bibfnamefont {A.~C.}\ \bibnamefont {Neto}},\ }\href@noop {} {\bibfield
  {journal} {\bibinfo  {journal} {Reviews of modern physics}\ }\textbf
  {\bibinfo {volume} {84}},\ \bibinfo {pages} {1067} (\bibinfo {year}
  {2012})}\BibitemShut {NoStop}%
\bibitem [{\citenamefont {Drut}\ and\ \citenamefont
  {L{\"a}hde}(2009)}]{drut2009graphene}%
  \BibitemOpen
  \bibfield  {author} {\bibinfo {author} {\bibfnamefont {J.~E.}\ \bibnamefont
  {Drut}}\ and\ \bibinfo {author} {\bibfnamefont {T.~A.}\ \bibnamefont
  {L{\"a}hde}},\ }\href@noop {} {\bibfield  {journal} {\bibinfo  {journal}
  {Physical review letters}\ }\textbf {\bibinfo {volume} {102}},\ \bibinfo
  {pages} {026802} (\bibinfo {year} {2009})}\BibitemShut {NoStop}%
\bibitem [{\citenamefont {Holstein}\ and\ \citenamefont
  {Primakoff}(1940)}]{holstein1940field}%
  \BibitemOpen
  \bibfield  {author} {\bibinfo {author} {\bibfnamefont {T.}~\bibnamefont
  {Holstein}}\ and\ \bibinfo {author} {\bibfnamefont {H.}~\bibnamefont
  {Primakoff}},\ }\href@noop {} {\bibfield  {journal} {\bibinfo  {journal}
  {Physical Review}\ }\textbf {\bibinfo {volume} {58}},\ \bibinfo {pages}
  {1098} (\bibinfo {year} {1940})}\BibitemShut {NoStop}%
\bibitem [{\citenamefont {Khomskii}(2010)}]{khomskii2010basic}%
  \BibitemOpen
  \bibfield  {author} {\bibinfo {author} {\bibfnamefont {D.~I.}\ \bibnamefont
  {Khomskii}},\ }\href@noop {} {\emph {\bibinfo {title} {Basic aspects of the
  quantum theory of solids: order and elementary excitations}}}\ (\bibinfo
  {publisher} {Cambridge university press},\ \bibinfo {year}
  {2010})\BibitemShut {NoStop}%
\bibitem [{Note1()}]{Note1}%
  \BibitemOpen
  \bibinfo {note} {We note that in contrast to the Fermi systems where the
  dynamical excitation matrix is identical to the Hamiltonian coefficient
  matrix, the dynamic matrix is distinct in the case of Bose systems~\cite
  {xiao2009theory}. Therefore care must be taken when calculating the magnon
  excitation spectrum.}\BibitemShut {Stop}%
\bibitem [{\citenamefont {Owerre}(2017)}]{owerre2017dirac}%
  \BibitemOpen
  \bibfield  {author} {\bibinfo {author} {\bibfnamefont {S.}~\bibnamefont
  {Owerre}},\ }\href@noop {} {\bibfield  {journal} {\bibinfo  {journal}
  {Scientific reports}\ }\textbf {\bibinfo {volume} {7}},\ \bibinfo {pages}
  {6931} (\bibinfo {year} {2017})}\BibitemShut {NoStop}%
\bibitem [{\citenamefont {Novoselov}(2011)}]{novoselov2011nobel}%
  \BibitemOpen
  \bibfield  {author} {\bibinfo {author} {\bibfnamefont {K.}~\bibnamefont
  {Novoselov}},\ }\href@noop {} {\bibfield  {journal} {\bibinfo  {journal}
  {Reviews of modern physics}\ }\textbf {\bibinfo {volume} {83}},\ \bibinfo
  {pages} {837} (\bibinfo {year} {2011})}\BibitemShut {NoStop}%
\bibitem [{\citenamefont {Neto}\ \emph {et~al.}(2009)\citenamefont {Neto},
  \citenamefont {Guinea}, \citenamefont {Peres}, \citenamefont {Novoselov},\
  and\ \citenamefont {Geim}}]{neto2009electronic}%
  \BibitemOpen
  \bibfield  {author} {\bibinfo {author} {\bibfnamefont {A.~C.}\ \bibnamefont
  {Neto}}, \bibinfo {author} {\bibfnamefont {F.}~\bibnamefont {Guinea}},
  \bibinfo {author} {\bibfnamefont {N.~M.}\ \bibnamefont {Peres}}, \bibinfo
  {author} {\bibfnamefont {K.~S.}\ \bibnamefont {Novoselov}}, \ and\ \bibinfo
  {author} {\bibfnamefont {A.~K.}\ \bibnamefont {Geim}},\ }\href@noop {}
  {\bibfield  {journal} {\bibinfo  {journal} {Reviews of modern physics}\
  }\textbf {\bibinfo {volume} {81}},\ \bibinfo {pages} {109} (\bibinfo {year}
  {2009})}\BibitemShut {NoStop}%
\bibitem [{\citenamefont {Fransson}\ \emph {et~al.}(2016)\citenamefont
  {Fransson}, \citenamefont {Black-Schaffer},\ and\ \citenamefont
  {Balatsky}}]{fransson2016magnon}%
  \BibitemOpen
  \bibfield  {author} {\bibinfo {author} {\bibfnamefont {J.}~\bibnamefont
  {Fransson}}, \bibinfo {author} {\bibfnamefont {A.~M.}\ \bibnamefont
  {Black-Schaffer}}, \ and\ \bibinfo {author} {\bibfnamefont {A.~V.}\
  \bibnamefont {Balatsky}},\ }\href@noop {} {\bibfield  {journal} {\bibinfo
  {journal} {Physical Review B}\ }\textbf {\bibinfo {volume} {94}},\ \bibinfo
  {pages} {075401} (\bibinfo {year} {2016})}\BibitemShut {NoStop}%
\bibitem [{\citenamefont {Owerre}(2016)}]{owerre2016first}%
  \BibitemOpen
  \bibfield  {author} {\bibinfo {author} {\bibfnamefont {S.}~\bibnamefont
  {Owerre}},\ }\href@noop {} {\bibfield  {journal} {\bibinfo  {journal}
  {Journal of Physics: Condensed Matter}\ }\textbf {\bibinfo {volume} {28}},\
  \bibinfo {pages} {386001} (\bibinfo {year} {2016})}\BibitemShut {NoStop}%
\bibitem [{\citenamefont {Pershoguba}\ \emph {et~al.}(2018)\citenamefont
  {Pershoguba}, \citenamefont {Banerjee}, \citenamefont {Lashley},
  \citenamefont {Park}, \citenamefont {{\AA}gren}, \citenamefont {Aeppli},\
  and\ \citenamefont {Balatsky}}]{pershoguba2018dirac}%
  \BibitemOpen
  \bibfield  {author} {\bibinfo {author} {\bibfnamefont {S.~S.}\ \bibnamefont
  {Pershoguba}}, \bibinfo {author} {\bibfnamefont {S.}~\bibnamefont
  {Banerjee}}, \bibinfo {author} {\bibfnamefont {J.}~\bibnamefont {Lashley}},
  \bibinfo {author} {\bibfnamefont {J.}~\bibnamefont {Park}}, \bibinfo {author}
  {\bibfnamefont {H.}~\bibnamefont {{\AA}gren}}, \bibinfo {author}
  {\bibfnamefont {G.}~\bibnamefont {Aeppli}}, \ and\ \bibinfo {author}
  {\bibfnamefont {A.~V.}\ \bibnamefont {Balatsky}},\ }\href@noop {} {\bibfield
  {journal} {\bibinfo  {journal} {Physical Review X}\ }\textbf {\bibinfo
  {volume} {8}},\ \bibinfo {pages} {011010} (\bibinfo {year}
  {2018})}\BibitemShut {NoStop}%
\bibitem [{\citenamefont {Onose}\ \emph {et~al.}(2010)\citenamefont {Onose},
  \citenamefont {Ideue}, \citenamefont {Katsura}, \citenamefont {Shiomi},
  \citenamefont {Nagaosa},\ and\ \citenamefont
  {Tokura}}]{onose2010observation}%
  \BibitemOpen
  \bibfield  {author} {\bibinfo {author} {\bibfnamefont {Y.}~\bibnamefont
  {Onose}}, \bibinfo {author} {\bibfnamefont {T.}~\bibnamefont {Ideue}},
  \bibinfo {author} {\bibfnamefont {H.}~\bibnamefont {Katsura}}, \bibinfo
  {author} {\bibfnamefont {Y.}~\bibnamefont {Shiomi}}, \bibinfo {author}
  {\bibfnamefont {N.}~\bibnamefont {Nagaosa}}, \ and\ \bibinfo {author}
  {\bibfnamefont {Y.}~\bibnamefont {Tokura}},\ }\href@noop {} {\bibfield
  {journal} {\bibinfo  {journal} {Science}\ }\textbf {\bibinfo {volume}
  {329}},\ \bibinfo {pages} {297} (\bibinfo {year} {2010})}\BibitemShut
  {NoStop}%
\bibitem [{\citenamefont {Coldea}\ \emph {et~al.}(2001)\citenamefont {Coldea},
  \citenamefont {Hayden}, \citenamefont {Aeppli}, \citenamefont {Perring},
  \citenamefont {Frost}, \citenamefont {Mason}, \citenamefont {Cheong},\ and\
  \citenamefont {Fisk}}]{coldea2001spin}%
  \BibitemOpen
  \bibfield  {author} {\bibinfo {author} {\bibfnamefont {R.}~\bibnamefont
  {Coldea}}, \bibinfo {author} {\bibfnamefont {S.}~\bibnamefont {Hayden}},
  \bibinfo {author} {\bibfnamefont {G.}~\bibnamefont {Aeppli}}, \bibinfo
  {author} {\bibfnamefont {T.}~\bibnamefont {Perring}}, \bibinfo {author}
  {\bibfnamefont {C.}~\bibnamefont {Frost}}, \bibinfo {author} {\bibfnamefont
  {T.}~\bibnamefont {Mason}}, \bibinfo {author} {\bibfnamefont {S.-W.}\
  \bibnamefont {Cheong}}, \ and\ \bibinfo {author} {\bibfnamefont
  {Z.}~\bibnamefont {Fisk}},\ }\href@noop {} {\bibfield  {journal} {\bibinfo
  {journal} {Physical Review Letters}\ }\textbf {\bibinfo {volume} {86}},\
  \bibinfo {pages} {5377} (\bibinfo {year} {2001})}\BibitemShut {NoStop}%
\bibitem [{\citenamefont {Li}\ \emph {et~al.}(2022)\citenamefont {Li},
  \citenamefont {Mai}, \citenamefont {Karaki}, \citenamefont {Jasper},
  \citenamefont {Garrity}, \citenamefont {Lyon}, \citenamefont {Shaw},
  \citenamefont {DeLazzer}, \citenamefont {Biacchi}, \citenamefont {Dally}
  \emph {et~al.}}]{li2022ring}%
  \BibitemOpen
  \bibfield  {author} {\bibinfo {author} {\bibfnamefont {Y.}~\bibnamefont
  {Li}}, \bibinfo {author} {\bibfnamefont {T.~T.}\ \bibnamefont {Mai}},
  \bibinfo {author} {\bibfnamefont {M.}~\bibnamefont {Karaki}}, \bibinfo
  {author} {\bibfnamefont {E.}~\bibnamefont {Jasper}}, \bibinfo {author}
  {\bibfnamefont {K.}~\bibnamefont {Garrity}}, \bibinfo {author} {\bibfnamefont
  {C.}~\bibnamefont {Lyon}}, \bibinfo {author} {\bibfnamefont {D.}~\bibnamefont
  {Shaw}}, \bibinfo {author} {\bibfnamefont {T.}~\bibnamefont {DeLazzer}},
  \bibinfo {author} {\bibfnamefont {A.}~\bibnamefont {Biacchi}}, \bibinfo
  {author} {\bibfnamefont {R.}~\bibnamefont {Dally}},  \emph {et~al.},\
  }\href@noop {} {\bibfield  {journal} {\bibinfo  {journal} {arXiv preprint
  arXiv:2212.05278}\ } (\bibinfo {year} {2022})}\BibitemShut {NoStop}%
\bibitem [{\citenamefont {Cookmeyer}\ \emph {et~al.}(2021)\citenamefont
  {Cookmeyer}, \citenamefont {Motruk},\ and\ \citenamefont
  {Moore}}]{cookmeyer2021four}%
  \BibitemOpen
  \bibfield  {author} {\bibinfo {author} {\bibfnamefont {T.}~\bibnamefont
  {Cookmeyer}}, \bibinfo {author} {\bibfnamefont {J.}~\bibnamefont {Motruk}}, \
  and\ \bibinfo {author} {\bibfnamefont {J.~E.}\ \bibnamefont {Moore}},\
  }\href@noop {} {\bibfield  {journal} {\bibinfo  {journal} {Physical Review
  Letters}\ }\textbf {\bibinfo {volume} {127}},\ \bibinfo {pages} {087201}
  (\bibinfo {year} {2021})}\BibitemShut {NoStop}%
\bibitem [{\citenamefont {Xiao}(2009)}]{xiao2009theory}%
  \BibitemOpen
  \bibfield  {author} {\bibinfo {author} {\bibfnamefont {M.-W.}\ \bibnamefont
  {Xiao}},\ }\href@noop {} {\bibfield  {journal} {\bibinfo  {journal} {arXiv
  preprint arXiv:0908.0787}\ } (\bibinfo {year} {2009})}\BibitemShut {NoStop}%
\end{thebibliography}%


\begin{thebibliography}{1}%
\makeatletter
\providecommand \@ifxundefined [1]{%
 \@ifx{#1\undefined}
}%
\providecommand \@ifnum [1]{%
 \ifnum #1\expandafter \@firstoftwo
 \else \expandafter \@secondoftwo
 \fi
}%
\providecommand \@ifx [1]{%
 \ifx #1\expandafter \@firstoftwo
 \else \expandafter \@secondoftwo
 \fi
}%
\providecommand \natexlab [1]{#1}%
\providecommand \enquote  [1]{``#1''}%
\providecommand \bibnamefont  [1]{#1}%
\providecommand \bibfnamefont [1]{#1}%
\providecommand \citenamefont [1]{#1}%
\providecommand \href@noop [0]{\@secondoftwo}%
\providecommand \href [0]{\begingroup \@sanitize@url \@href}%
\providecommand \@href[1]{\@@startlink{#1}\@@href}%
\providecommand \@@href[1]{\endgroup#1\@@endlink}%
\providecommand \@sanitize@url [0]{\catcode `\\12\catcode `\$12\catcode
  `\&12\catcode `\#12\catcode `\^12\catcode `\_12\catcode `\%12\relax}%
\providecommand \@@startlink[1]{}%
\providecommand \@@endlink[0]{}%
\providecommand \url  [0]{\begingroup\@sanitize@url \@url }%
\providecommand \@url [1]{\endgroup\@href {#1}{\urlprefix }}%
\providecommand \urlprefix  [0]{URL }%
\providecommand \Eprint [0]{\href }%
\providecommand \doibase [0]{http://dx.doi.org/}%
\providecommand \selectlanguage [0]{\@gobble}%
\providecommand \bibinfo  [0]{\@secondoftwo}%
\providecommand \bibfield  [0]{\@secondoftwo}%
\providecommand \translation [1]{[#1]}%
\providecommand \BibitemOpen [0]{}%
\providecommand \bibitemStop [0]{}%
\providecommand \bibitemNoStop [0]{.\EOS\space}%
\providecommand \EOS [0]{\spacefactor3000\relax}%
\providecommand \BibitemShut  [1]{\csname bibitem#1\endcsname}%
\let\auto@bib@innerbib\@empty
\bibitem [{\citenamefont {Xiang}\ \emph {et~al.}(2013)\citenamefont {Xiang},
  \citenamefont {Lee}, \citenamefont {Koo}, \citenamefont {Gong},\ and\
  \citenamefont {Whangbo}}]{xiang2013magnetic}%
  \BibitemOpen
  \bibfield  {author} {\bibinfo {author} {\bibfnamefont {H.}~\bibnamefont
  {Xiang}}, \bibinfo {author} {\bibfnamefont {C.}~\bibnamefont {Lee}}, \bibinfo
  {author} {\bibfnamefont {H.-J.}\ \bibnamefont {Koo}}, \bibinfo {author}
  {\bibfnamefont {X.}~\bibnamefont {Gong}}, \ and\ \bibinfo {author}
  {\bibfnamefont {M.-H.}\ \bibnamefont {Whangbo}},\ }\href@noop {} {\bibfield
  {journal} {\bibinfo  {journal} {Dalton Transactions}\ }\textbf {\bibinfo
  {volume} {42}},\ \bibinfo {pages} {823} (\bibinfo {year} {2013})}\BibitemShut
  {NoStop}%
\end{thebibliography}%

\end{document}


\title{--Supplementary Materials--\\Coexistence of Dirac Fermions and Magnons in a Layered Two-Dimensional Semiquinoid  Metal-Organic Framework}

\author{Christopher Lane}
\email{laneca@lanl.gov}
\affiliation{Theoretical Division, Los Alamos National Laboratory, Los Alamos, New Mexico 87545, USA}
\affiliation{Center for Integrated Nanotechnologies, Los Alamos National Laboratory, Los Alamos, New Mexico 87545, USA}

\author{Yixuan Huang}
\affiliation{Theoretical Division, Los Alamos National Laboratory, Los Alamos, New Mexico 87545, USA}
\affiliation{Center for Integrated Nanotechnologies, Los Alamos National Laboratory, Los Alamos, New Mexico 87545, USA}

\author{Lin Hou}
\affiliation{Theoretical Division, Los Alamos National Laboratory, Los Alamos, New Mexico 87545, USA}

\author{Jian-Xin Zhu}
\affiliation{Theoretical Division, Los Alamos National Laboratory, Los Alamos, New Mexico 87545, USA}
\affiliation{Center for Integrated Nanotechnologies, Los Alamos National Laboratory, Los Alamos, New Mexico 87545, USA}

\date{\today} 

\pacs{}

\maketitle

\tableofcontents

\clearpage
\section{Unrelaxed Crystal Results}
In this section, we present the exchange parameters obtained using the experimental crystal structure without structural relaxation.

\subsection{Exchange Parameters}
Table~\ref{tab:unrelax_exchange} presents the first-, second-, and third-nearest neighbor exchange parameters, along with the spin magnitude for all 3$d$ transition metals in the experimental crystal structure [M$_{2}$L$_{3}$]$^{2-}$ MOF. 

\begin{table}[h!]
\begin{ruledtabular}
\begin{tabular}{c|c|c|c|c|c}
 & $J$ (meV) & $J^\prime$ (meV) & $J^{\prime\prime}$ (meV) & $J_{\perp}$ (meV) & $S$ \\
\hline\hline
Ti & $-$       & $-$        & $-$         & $-$          & 0.0 \\
V  & -7.8704   & 1.2844     & 2.4126      & -1.1097E-02  & 1   \\
Cr & 9.6971    & 9.91E-02   & 1.11E-02    & -2.9555E-04  & 1.5 \\
Mn & -21.4668  & 3.90E-02   & -4.9882     & -3.2499E-04  & 1.5 \\
Fe & 1.8228    & 6.55E-03   & 9.01E-03    & 8.52187E-04  & 2   \\
Co & -21.6576  & 3.0208     & -2.3148     & -3.6692E-02  & 1.5 \\
Ni & -21.9392  & 3.9891     & -17.3706    & -0.88693     & 1   \\
Cu & -24.9644  & 0.8441     & -0.3804     & -2.7028      & 0.5 \\
Zn & $-$       & $-$        & $-$         & $-$          & 0.0 \\
\end{tabular}
\end{ruledtabular}
\caption{Intra- and inter-planar exchange parameters and spin magnitude in the unrelaxed metal-semiquinoid framework (H$_2$NMe$_2$)$_2$M$_2$(Cl$_2$dhbq)$_3$ for all $3d$ transition-metals.}
\label{tab:unrelax_exchange}
\end{table}

\clearpage
\section{Relaxed Crystal Results}
In this section, we present the exchange parameters, band dispersions and magnon dispersions obtained using the relaxed crystal structure, whereas all figures in the main text are based on the experimental structure.

\subsection{Exchange Parameters}
\begin{figure*}[ht!]
\includegraphics[width=0.95\textwidth]{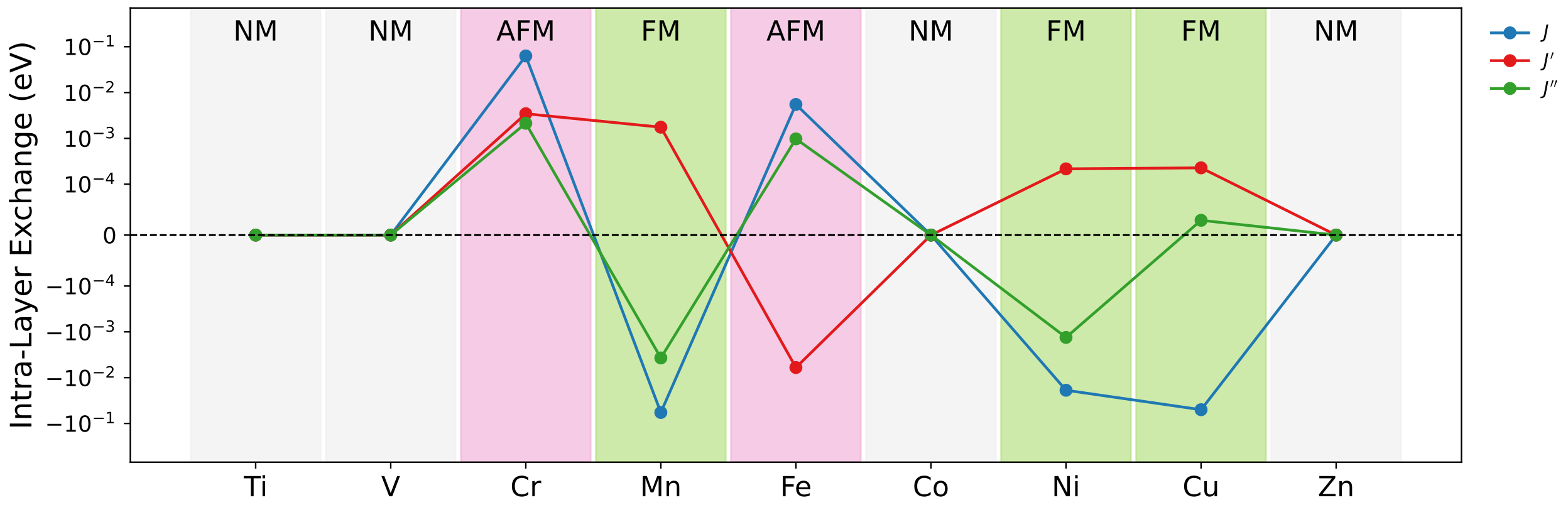}
\caption{Plot of exchange parameters in energy units as a function of transition-metal.}
\label{fig:relaxed_mag_exchange}
\end{figure*}

Figure~\ref{fig:relaxed_mag_exchange} presents the first-, second-, and third-nearest neighbor exchange parameters for all 3$d$ transition metals in the [M$_{2}$L$_{3}$]$^{2-}$ MOF. Table~\ref{tab:relaxed_exchange_parameters} gives the values along with the spin magnitude. Ti, and Zn V, and Co display negligible and zero magnetic polarization on the atomic sites, respectively, yielding no exchange parameters. For transition-metals to the left of Co on the periodic table, $J$ and $J^{\prime}$ alternate between positive and negative values, whereas for Ni, and Cu $J$ and $J^{\prime}$ are consistently negative and positive, respectively. The next-nearest neighbor exchange coupling is found to be finite, despite the $\sim$13.31~\AA~ updated from relaxed structures separation, and positive for all magnetic ions except for Fe. 

The magnetic ground state of [M$_{2}$L$_{3}$]$^{2-}$ MOF for each transition metal is determined by solving for the ground state of the next-next-nearest neighbor Heisenberg model within the DMRG. The resulting magnetic ground state for each transition metal is indicated in Fig.~\ref{fig:relaxed_mag_exchange}. We find every transition-metal ion to be in the low spin state, with spin quantum numbers $s = 1$ for Cr and Mn, and $s = 1/2$ for Fe, Ni, and Cu. As atomic number increase from 22, the ground state alternates between ferromagnetic (FM) and N\'{e}el-type antiferromagnetic (AFM) ordering up through Fe. For Ni and Cu the ground state settles into a robust FM order.

\begin{table}[h!]
\begin{ruledtabular}
\begin{tabular}{c|c|c|c|c}
 &$J$ (meV)&$J^\prime$ (meV)&$J^{\prime\prime}$ (meV)&$S$\\
\hline\hline
Ti & $-$         & $-$           & $-$           & $0$     \\
V  & $-$         & $-$           & $-$           & $0$     \\
Cr & 0.06291854  & 0.00344551    & 0.00213417    & $1$     \\
Mn & $-0.0575873$& 0.00175138    & $-0.0037165$  & $1$     \\
Fe & 0.00549679  & $-0.0060195$  & 0.00097367    & $1/2$   \\
Co & $-$         & $-$           & $-$           & $0$     \\
Ni & $-0.0188515$& 0.00021595    & $-0.001325$   & $1/2$   \\
Cu & $-0.050133$ & 0.00022706    & 2.9032E-05    & $1/2$   \\
Zn & $-$         & $-$           & $-$           & $0$     \\
\end{tabular}
\end{ruledtabular}
\caption{Intra- and inter-planar exchange parameters and spin magnitude in the relaxed metal-semiquinoid framework (H$_2$NMe$_2$)$_2$M$_2$(Cl$_2$dhbq)$_3$ for all $3d$ transition-metals.}
\label{tab:relaxed_exchange_parameters}
\end{table}

\clearpage
\subsection{Band Dispersions}
\begin{figure*}[ht!]
\includegraphics[width=.99\textwidth]{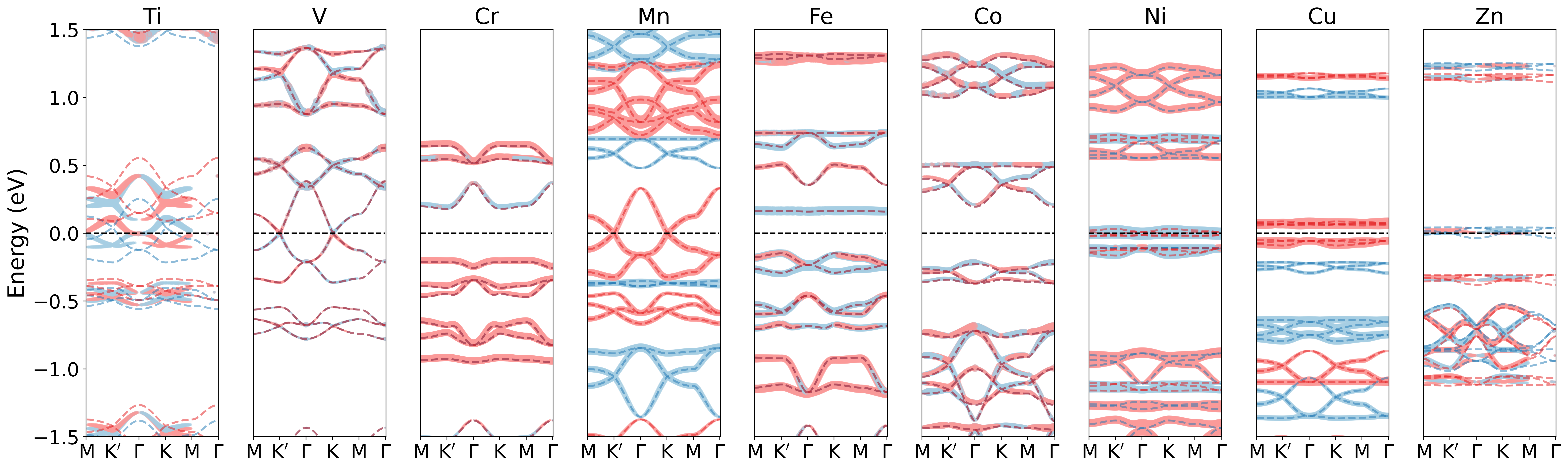}
\caption{(color online) Spin-resolved electronic band dispersions with (dots) and without (dashed line) spin-orbit coupling of the metal-semiquinoid framework (H$_2$NMe$_2$)$_2$M$_2$(Cl$_2$dhbq)$_3$ for various $3d$ transition-metals. The sizes of the dots are proportional to the spin-polarization $\braket{S_{z}}$, with the spin-up (down) states indicated in red (blue) color. } 
\label{fig:relaxed_bands}
\end{figure*}

Figure~\ref{fig:relaxed_bands} shows the spin-resolved electronic band dispersion for each [M$_{2}$L$_{3}$]$^{2-}$ in its associated magnetic ground state. A rich variety of electronic structures are exhibited showing a sensitive dependence on magnetic ground state and $d$-electron count. Specifically, Cr and Fe, Co display an AFM and a NM insulating state, respectively, with highly localized carriers as demonstrated by the extremely flat bands at the valence band edges. 

Mn and V host fully spin-polarized and spin unpolarized Dirac fermions at the Fermi level that are clearly separated from other bands in the system by $\sim 300$ meV. Interestingly, the bandwidth and Fermi velocity of the linearly dispersing states of the Ni and Zn-based compound are significantly reduced compared to those in Mn, indicating an enhancement in the electron–electron interaction strength. In contrast, FM Ni is a metal and FM Cu is a narrow in-direct band gap ($\sim 25$ meV) FM semiconductor. Finally, under electron doping Ti is in close proximity to the elusive nearly nonmagnetic spin–valley coupled Dirac semimetal phase. 

\clearpage
\subsection{Magnon Dispersions}
\begin{figure*}[ht!]
\includegraphics[width=.99\textwidth]{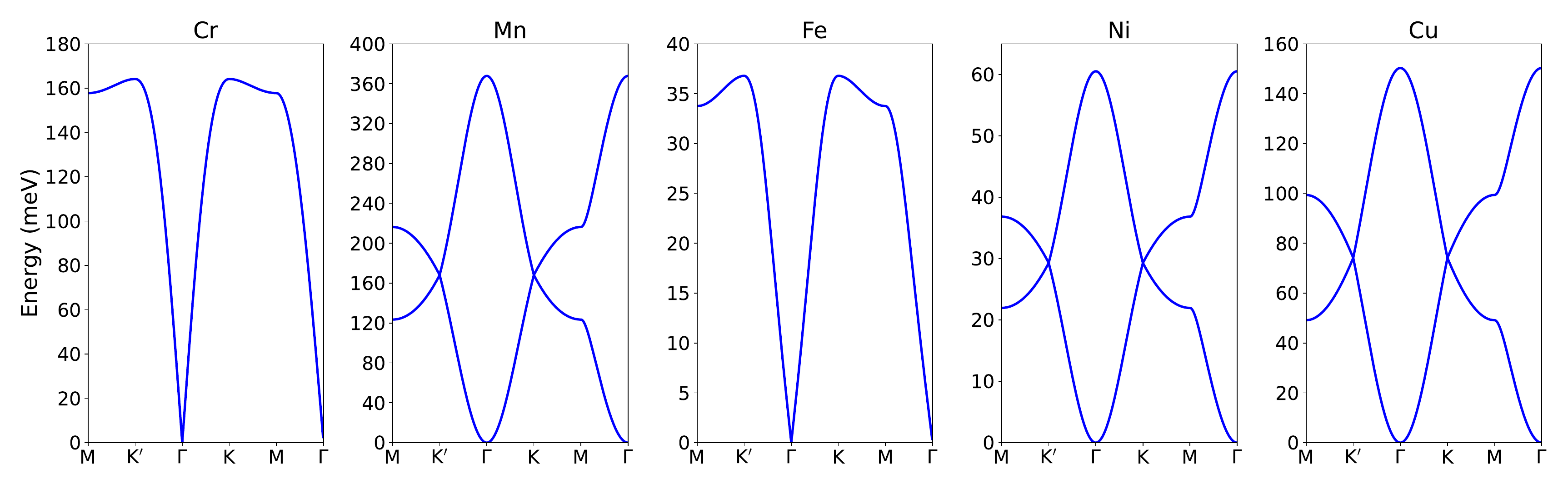}
\caption{(color online) Magnon excitation spectrum of the metal-semiquinoid framework (H$_2$NMe$_2$)$_2$M$_2$(Cl$_2$dhbq)$_3$ for various $3d$ transition-metals.} 
\label{fig:relaxed_magnon_dispersion}
\end{figure*}

Figure~\ref{fig:relaxed_magnon_dispersion} presents the magnon dispersion for each 3$d$ translation metal. In AFM Cr and Fe based semiquinoid metal-organic frameworks a single magnon band is found that follows a linear dispersion to zero at $\Gamma$, as is typical for antiferromagents. FM Mn, Ni, and Cu based compounds display clear Dirac crossings at K (K$^\prime$), along with the standard quadratic band dispersion at $\Gamma$. Here, we find the [M$_{2}$L$_{3}$]$^{2-}$ MOF composed of Mn to display the coexistence of Direc fermions and magnons which allows for the intertwining of non-trivial fermionic and bosonic band topologies.  

\clearpage
\section{Mapping Total Energies to Magnetic Exchange Parameters}
\begin{figure}[ht!]
\includegraphics[width=.40\textwidth]{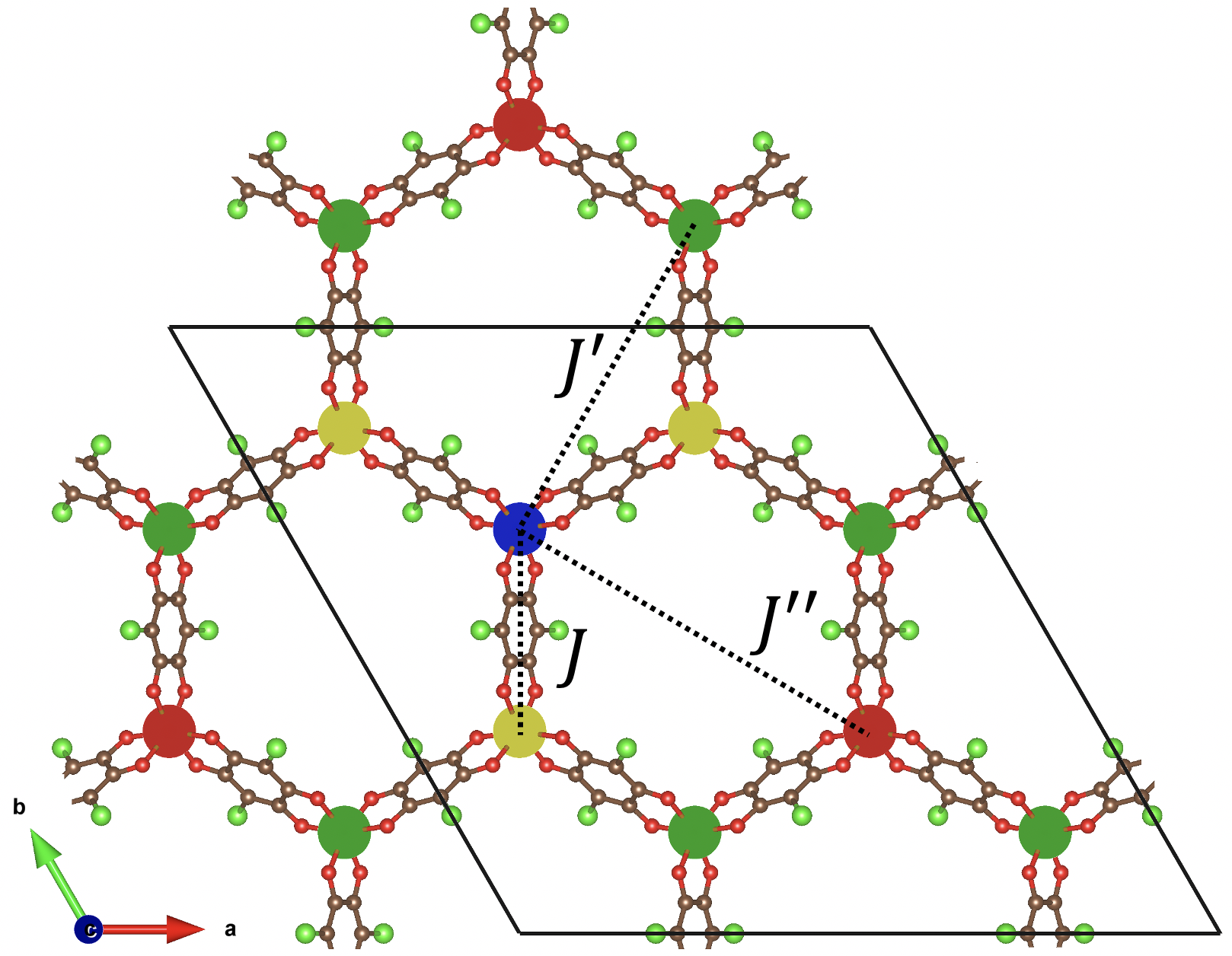}
\caption{(color online) Schematic of the central magnetic ion (blue dot), nearest-neighbors (yellow dot), next-nearest-neighbors (green dot), and next-next-nearest-neighbors (red dot)
$J$, $J^{\prime}$, and $J^{\prime\prime}$ exchange parameters overlaid on the honeycomb lattice of transition-metal sites within a $2\times 2$ super cell of the [M$_{2}$L$_{3}$]$^{2-}$ metal-organic framework. The black lines mark the unit cell. } 
\label{fig:magnetic_exchange_schematic}
\end{figure}
In order to determine the strength of the exchange coupling between neighboring transition-metals in the [M$_{2}$L$_{3}$]$^{2-}$ MOF, we map the total energies of the various spin configurations onto those of the third-nearest-neighbor Heisenberg Hamiltonian in the mean-field approximation~\cite{xiang2013magnetic}. Figure~\ref{fig:magnetic_exchange_schematic} shows a $2\times 2$ super cell of the [M$_{2}$L$_{3}$]$^{2-}$ MOF with the color of the transition-metal sites marking the central magnetic ion (blue dot), nearest-neighbors (yellow dot), next-nearest-neighbors (green dot), and next-next-nearest-neighbors (red dot). For this crystal lattice, we can write the Heisenberg spin Hamiltonian as,
\begin{align}
\mathcal{H}=&\sum_{\braket{i<j}}J_{ij}S_{i}S_{j}+\sum_{\braket{\braket{i<j}}}J_{ij}^{\prime}S_{i}S_{j}+\sum_{\braket{\braket{\braket{i<j}}}}J_{ij}^{\prime\prime}S_{i}S_{j},\label{eq:Hnn}
\end{align}
where $i(j)$ indexes the magnetic ion lattice positions, $S_{i}$ is the local magnetic moment on lattice site $i$, $J_{ij}$, $J_{ij}^{\prime}$, and $J_{ij}^{\prime\prime}$ denote the 
nearest-neighbor, next-nearest-neighbor, and third-nearest-neighbor exchange interaction strength between sites $i$ and $j$ such that 
 $J_{ij}^{(\prime,\prime\prime)}=J_{ji}^{(\prime,\prime\prime)}$. Furthermore, the number of magnetic configurations needed to fully determine the exchange parameters is reduced by assuming $J_{ij}$ to be bond independent. This allows us to re-label the various terms in the Hamiltonian by nearest-neighbor rings $a$, $b$, $c$, and $d$, yielding,
\begin{align}
\mathcal{H}=&3JS_{a}S_{b}+6J^{\prime}S_{a}S_{c}+3J^{\prime\prime}S_{a}S_{d},
\end{align}
where the leading integer coefficient is the coordination number for each ring of neighbors. 

Since the local magnetic moment on each site can be either $+$ or $-$, there are $2^{4}=16$ unique spin configurations that may be realized. Each configuration $\gamma$ is given by the direction of each moment for each successive nearest-neighbor ring, e.g., $+-+-$ meaning $\braket{S_{a}},\braket{S_{c}}>0$ and $\braket{S_{b}},\braket{S_{d}}<0$. Additionally, since the total energy of configurations $\gamma$ and $-\gamma$ are degenerate, the number of {\it ab intio} calculations is reduced by half. Finally, by summing configurations $\gamma$ and $-\gamma$ and using the resulting eight equations to solve for $J$, $J^{\prime}$, and $J^{\prime\prime}$, we find 
{\small
\begin{align}
J=\frac{(E^1+E^4+E^5+E^6)-(E^2+E^3+E^7+E^8)}{16\cdot 3\braket{S}^{2}},\\
J^{\prime}=\frac{(E^1+E^3+E^5+E^7)-(E^2+E^4+E^6+E^8)}{16\cdot 6\braket{S}^{2}},\\
J^{\prime\prime}=\frac{(E^1+E^3+E^4+E^8)-(E^2+E^5+E^6+E^7)}{16\cdot 3\braket{S}^{2}},
\end{align}
}
where we have defined,
\begin{align}
E^{1}=E^{++++}+E^{----}, ~&~ E^{5}=E^{+++-}+E^{---+},\nonumber\\
E^{2}=E^{-+++}+E^{+---}, ~&~ E^{6}=E^{++--}+E^{--++},\nonumber\\
E^{3}=E^{+-++}+E^{-+--}, ~&~ E^{7}=E^{+-+-}+E^{-+-+},\nonumber\\
E^{4}=E^{++-+}+E^{--+-}, ~&~ E^{8}=E^{+--+}+E^{-++-}.\nonumber
\end{align}

In addition to the intra-layer exchange interactions, we calculated the inter-layer nearest-neighbor exchange coupling to gauge the degree of magnetic isolation of each layer. By considering parallel and anti-parallel spin configurations between the layers in the $1\times 1\times 2$ super cell, we find 
\begin{align}
J_{\perp}=\frac{ \left( E^{++}+E^{--} \right)- \left( E^{+-}+E^{-+} \right)}{4\cdot 2 \braket{S}^{2}}.
\end{align}

\clearpage
\section{Derivation of Magnon Dispersions}
The Dirac dispersion of magnons can be studied in a semi-classical magnon model. The effective spin interactions are described by the Heisenberg Hamiltonian $H=\sum _{\left \{ i,j \right \}} J_{ij}\mathbf{S}_{i}\cdot \mathbf{S}_{j}$, where we include interactions between nearest-neighbor $J$, next-nearest-neighbor $J^{\prime}$, and next-next-nearest-neighbor $J^{\prime\prime}$.

\subsection{The FM State}
\begin{figure}[!b]
\centering
\includegraphics[width=0.4\columnwidth]{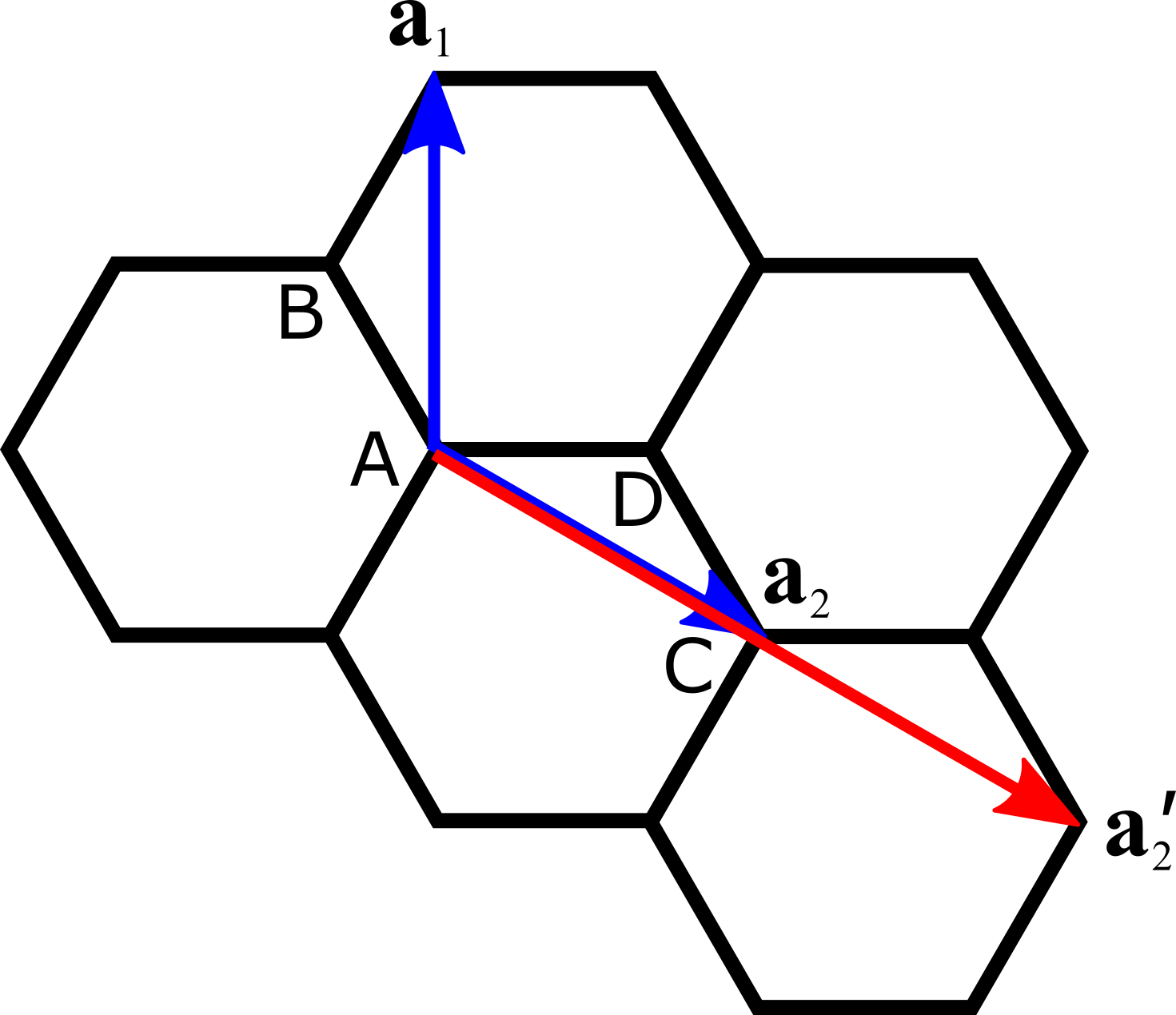}
\caption{Schematics of the honeycomb lattice with sub-lattice sites labeled as $A$, $B$, $C$, $D$, and lattice translation vectors $\mathbf{a}_{1},\mathbf{a}_{2}$ for the FM and N\'eel AFM states, and $\mathbf{a}_{1},\mathbf{a}'_{2}$ for the Zig-zag AFM state.}
\label{neighbor_vector}
\end{figure}

For ferromagnetic (FM) honeycomb lattice ground state, we label spins $A$ and $B$ within each unit cell [see Fig.~\ref{neighbor_vector}] and use the Holstein-Primakoff representation to recast the spin operators in terms of boson operators:
\begin{equation}
\begin{aligned}
\label{eqS_AFM1}
S_{i}^{+}=\sqrt{2S-a_{i}^{\dag}a_{i}}a_{i}, 
\ \
S_{i}^{-}=a_{i}^{\dag }\sqrt{2S-a_{i}^{\dag }a_{i}},
\ \ \mbox{and} \ \
S_{i}^{z}=S-a_{i}^{\dag }a_{i},
\end{aligned}
\end{equation}
for $i= A,B$. In the reciprocal space the boson operators are defined as $a_{A\mathbf{k}}=\frac{1}{^{\sqrt{N/2}}}\sum\nolimits_{i\in A}e^{-i\mathbf{k}\cdot \mathbf{r}_{i}}a_{i}$ and $a_{B\mathbf{k}}=\frac{1}{^{\sqrt{N/2}}}\sum\nolimits_{i\in B}e^{-i\mathbf{k}\cdot \mathbf{r}_{i}}b_{i} $. Using the basis of $\Phi _{\mathbf{k}}^{\dagger }=(a_{A\mathbf{k}}^{\dagger },a_{B\mathbf{k}}^{\dagger })$, the Hamiltonian is given by
\begin{widetext}
\begin{align}
\label{eq_FM}
H_{\text{FM}}(\mathbf{k})=S\sum _{\mathbf{k}}\Phi _{\mathbf{k}}^{\dagger } \begin{bmatrix}
\chi_{\text{FM}}(\mathbf{k}) & \phi(\mathbf{k}) \\
\phi ^{*}(\mathbf{k}) & \chi_{\text{FM}}(\mathbf{k})
\end{bmatrix} \Phi _{\mathbf{k}},
\end{align}
\end{widetext}
where 
\begin{align*}
&\chi_{\text{FM}}(\mathbf{k})=-3J+(2\cos(\mathbf{k} \cdot \mathbf{a}_{1})+2\cos(\mathbf{k} \cdot \mathbf{a}_{2})+2\cos[\mathbf{k} \cdot (\mathbf{a}_{1}+\mathbf{a}_{2})]-6)J^{\prime}-3J^{\prime\prime},\\
&\phi (\mathbf{k})=(1+e^{i\mathbf{k} \cdot \mathbf{a}_{1}}+e^{-i\mathbf{k} \cdot \mathbf{a}_{2}})J+(e^{i\mathbf{k} \cdot (\mathbf{a}_{1}+\mathbf{a}_{2})}+e^{i\mathbf{k} \cdot (\mathbf{a}_{1}-\mathbf{a}_{2})}+e^{-i\mathbf{k} \cdot (\mathbf{a}_{1}+\mathbf{a}_{2})})J^{\prime\prime}, 
\end{align*}
and $\mathbf{a}_{i}$ is a lattice vector defined as $\mathbf{a}_{1}=(3/2,-\sqrt{3}/2)$ and $\mathbf{a}_{2}=(0,\sqrt{3})$ in the Cartesian coordinates, see Fig.~\ref{neighbor_vector}. The eigenvalues of $H_{FM}$ yield the energy dispersion of the ferromagnetic magnons: $\varepsilon^{\pm}_{\text{FM}}(\mathbf{k})=S (\chi_{\text{FM}}(\mathbf{k})\pm |\phi (\mathbf{k})|)$. 

By inspection, we find band degeneracies in the dispersion at $\mathbf{K}(\mathbf{K}^{\prime})$ points in the Brillouin zone. That is, if we expand $\varepsilon^{\pm}_{\text{FM}}(\mathbf{k})$ to first order in the neighborhood of $\mathbf{K}=(0,\frac{4\pi }{3\sqrt{3}})$,
\begin{widetext}
\begin{align}
\label{eqS_FM1}
\varepsilon^{\pm}_{\text{FM}}(\mathbf{K}+d\mathbf{k}) &=S(\chi_{\text{FM}}(\mathbf{K}+d\mathbf{k})\pm |\phi (\mathbf{K}+d\mathbf{k})|),\\
&=-S(3J+9J^{\prime}+3J^{\prime\prime}) \pm S(3J/2-3J^{\prime\prime})|d\mathbf{k}|+O(|d\mathbf{k}|^{2})
\end{align}
\end{widetext}
where $d\mathbf{k}=(dk_{x},dk_{y})$, we find the dispersion to linearly depend on $d\mathbf{k}$. Interestingly, when $J=2J^{\prime\prime}$ the linear terms disappears and the energy of magnons at $\mathbf{K}(\mathbf{K}^{\prime})$ is $\varepsilon ^{\pm}_{FM }(\mathbf{K})=-S(3J+9J^{\prime}+3J^{\prime\prime})$. In the present work, all FM cases studied display a linear dispersion.

\subsection{The N\'eel AFM State}
The N\'eel antiferromagnetic (AFM) state has two opposite spins in a unit cell, which we label as $A$ and $B$. The Holstein-Primakoff transformations for spins $A$ and $B$ are defined as:
\begin{equation}
\begin{aligned}
\label{eqS_AFM1}
&S_{i}^{+}=\sqrt{2S-a_{i}^{\dag }a_{i}}a_{i},
\ \ 
S_{i}^{-}=a_{i}^{\dag }\sqrt{2S-a_{i}^{\dag }a_{i}},
\ \ \mbox{and} \ \
S_{i}^{z}=S-a_{i}^{\dag }a_{i},\\
&S_{j}^{+}=b_{j}^{\dag }\sqrt{2S-b_{j}^{\dag }b_{j}},
\ \
S_{j}^{-}=\sqrt{2S-b_{j}^{\dag }b_{j}}~b_{j},
\ \ \mbox{and} \ \
S_{j}^{z}=-S+b_{j}^{\dag }b_{j},
\end{aligned}
\end{equation}
where boson operator $a^{\dag }_{i}$ ($b^{\dag }_{j}$) acts on $A$ $(B)$ spins. Fourier transforming the N\'eel AFM Heisenberg Hamiltonian to momentum space, yields: 
\begin{widetext}
\begin{align}
\label{eq_AFM}
H_{\text{N\'eel}}(\mathbf{k})=S\sum _{\mathbf{k}}\Psi _{\mathbf{k}}^{\dagger } \begin{bmatrix}
\chi_{\text{N\'eel}}(\mathbf{k}) & \phi(\mathbf{k}) \\
\phi ^{*}(\mathbf{k}) & \chi_{\text{N\'eel}}(\mathbf{k})
\end{bmatrix} \Psi _{\mathbf{k}}.
\end{align}
\end{widetext}
with $\Psi  _{\mathbf{k}}^{\dagger }=(a_{A\mathbf{k}}^{\dagger },b_{B-\mathbf{k}})$, 
\begin{align*}
\chi_{\text{N\'eel}}(\mathbf{k})=3J+(2\cos(\mathbf{k} \cdot \mathbf{a}_{1})+2\cos(\mathbf{k} \cdot \mathbf{a}_{2})+2\cos[\mathbf{k} \cdot (\mathbf{a}_{1}+\mathbf{a}_{2})]-6)J^{\prime}+3J^{\prime\prime},
\end{align*}
and $\phi (\mathbf{k})$ is the same as in the FM case. To obtain the magnon dispersion, we use the Bogoliubov transformation 
\begin{equation}
\begin{aligned}
\label{eqS_AFM2}
\alpha _{\mathbf{k}}=u _{\mathbf{k}} a_{A\mathbf{k}}-v_{\mathbf{k}}b_{B-\mathbf{k}}^{\dagger }, \\
\beta  _{\mathbf{k}}=u _{\mathbf{k}} b_{B-\mathbf{k}}-v_{\mathbf{k}}a_{A\mathbf{k}}^{\dagger },
\end{aligned}
\end{equation}
to diagonalize the Hamiltonian by basis transformation. We note that the quasi-particle excitations $\alpha _{\mathbf{k}}$ and $\beta  _{\mathbf{k}}$ obey bosonic commutation relation when $|u _{\mathbf{k}}|^{2}-|v _{\mathbf{k}}|^{2}=1$. The Hamiltonian is diagonal when 
\[
2E_{\text{N\'eel}}(\mathbf{k})u_{\mathbf{k}}v_{\mathbf{k}}+\phi ^{*} (\mathbf{k})u_{\mathbf{k}}^{2}+\phi  (\mathbf{k})v_{\mathbf{k}}^{2}=0.
\]
is satisfied. By examining the imaginary part of this expression, we find $\phi (\mathbf{k}) ,u_{\mathbf{k}}, v_{\mathbf{k}}$ has the following form
\begin{equation}
\begin{aligned}
\label{eqS_AFM3}
\phi (\mathbf{k})&=\left |\phi (\mathbf{k}) \right |e^{i\eta }, \\
u_{\mathbf{k}}&=\left | u _{\mathbf{k}} \right |e^{i(\eta /2 +\pi) }, \\
v _{\mathbf{k}}&=\left | v _{\mathbf{k}} \right |e^{-i\eta /2},
\end{aligned}
\end{equation}
where $\eta $ is a phase parameter that is fixed via $\phi (\mathbf{k})$ in Eq.~\ref{eqS_AFM3}. Now combining
\[
2\chi_{\text{N\'eel}}(\mathbf{k})|u_{\mathbf{k}}||v_{\mathbf{k}}|+|\phi  (\mathbf{k})||u_{\mathbf{k}}|^{2}+|\phi  (\mathbf{k})||v_{\mathbf{k}}|^{2}=0, 
\]
with the bosonic commutation relation $|u _{\mathbf{k}}|^{2}-|v _{\mathbf{k}}|^{2}=1$ we solve for $|u _{\mathbf{k}}|$ and $|v _{\mathbf{k}}|$. This finally yields the magnon energy dispersion:
\begin{align}
\label{eqS_AFM4}
\varepsilon_{\text{N\'eel}} (\mathbf{k})=S\sqrt{\chi_{\text{N\'eel}}^{2}(\mathbf{k})- |\phi (\mathbf{k})|^{2}}.
\end{align}

If we expand the magnon dispersion in a neighborhood around $\Gamma $
\begin{align}
\label{eqS_AFM5}
\varepsilon_{\text{N\'eel}}(d\mathbf{k})=S\sqrt{\frac{9}{2}J^{2}+\frac{45}{2}JJ^{\prime\prime}+18J^{\prime\prime 2}-27(J+J^{\prime\prime})J^{\prime}} |d\mathbf{k}|+O(|d\mathbf{k}|^{2}),
\end{align}
the leading term is linear. We find the dispersion to be linear for all N\'eel AFM cases studied.

\subsection{The Zig-zag AFM State}
For the zig-zag AFM in a honeycomb lattice we assign $A$, $B$, $C$, and $D$ spins in an enlarged $1\times 2$ unit cell with $A$ and $B$ spins pointing up while $C$ and $D$ spins pointing down, as shown in Fig.~\ref{neighbor_vector}. Similar to the Holstein-Primakoff transformation for N\'eel AFM state, the Heisenberg Hamiltonian is written as 
\begin{align}
\label{eqS_stripe1}
H_{\text{zig-zag}}(\mathbf{k})&=S\sum _{\mathbf{k}}\Lambda _{\mathbf{k}}^{\dagger } [H_{\text{zig-zag}}]_{\mathbf{k}} \Lambda _{\mathbf{k}}, \\
[H_{\text{zig-zag}}]_{\mathbf{k}}&=\begin{bmatrix}
\chi_{1} & \phi _{1} &\phi_{2}  & \phi _{3} \\ 
\phi _{1}^{*} & \chi_{1} & \phi _{4} &\phi _{2} \\ 
\phi_{2}^{*} & \phi _{4}^{*} & \chi_{1} & \phi _{1}\\ 
\phi _{3}^{*} & \phi_{2}^{*} & \phi _{1}^{*} & \chi_{1}
\end{bmatrix},
\end{align}
in the basis of $\Lambda_{\mathbf{k}}^{\dagger }=(a_{A\mathbf{k}}^{\dagger },a_{B\mathbf{k}}^{\dagger },b_{C-\mathbf{k}},b_{D-\mathbf{k}})$, with the various matrix elements defined as:
\begin{align*}
&\chi_{1}=-J+[2\cos(\sqrt{3}k_{y})+2]J^{\prime}+3J^{\prime\prime},\\
&\phi_{1}=(1+e^{-i\sqrt{3}k_{y}})J,\\ 
&\phi_{2}=(e^{i\sqrt{3}k_{y}}+1+e^{-i3k_{x}}+e^{-i(3k_{x}-\sqrt{3}k_{y}})J^{\prime}, \\
&\phi_{3}=J+(e^{i\sqrt{3}k_{y}}+e^{-i\sqrt{3}k_{y}}+e^{-i3k_{x}})J^{\prime\prime},\\
&\phi_{4}=e^{-i(3k_{x}-\sqrt{3}k_{y})}J+(e^{i\sqrt{3}k_{y}}+e^{i(-3k_{x}+2\sqrt{3}k_{y})}+e^{-i3k_{x}})J^{\prime\prime}. 
\end{align*}
Using the Bogoliubov transformation, the magnon energy dispersion is thus:
\begin{align}\label{eqS_stripe2}
\varepsilon^{\pm}_{\text{zig-zag}}(\mathbf{k})=&S\sqrt{\chi_{1}^{2}+\left|\phi_{1}\right|^{2}-\left|\phi_{2}\right|^{2}-\frac{1}{2}\left|\phi_{3}\right|^{2}-\frac{1}{2}\left|\phi_{4}\right|^{2} \pm \sqrt{F(\chi_{1},\phi_{1},\phi_{2},\phi_{3},\phi_{4})}}, \\
\end{align}
where
\begin{align}
F(\chi_{1},\phi_{1},\phi_{2},\phi_{3},\phi_{4}) =&4\chi_{1}^{2}\left|\phi_{1}\right|^{2}-2\chi_{1}\phi_{1}\phi_{2}^{*}\phi_{4}-2\chi_{1}\phi_{1}^{*}\phi_{2}^{*}\phi_{3}-2\chi_{1}\phi_{1}^{*}\phi_{2}\phi_{4}^{*}-2\chi_{1}\phi_{1}\phi_{2}\phi_{3}^{*}+\phi_{1}^{2}\phi_{3}^{*}\phi_{4} \nonumber\\
&+\phi_{2}^{*2}\phi_{3}\phi_{4}+\phi_{1}^{*2}\phi_{3}\phi_{4}^{*}+\phi_{2}^{2}\phi_{3}^{*}\phi_{4}^{*}-\left|\phi_{1}\right|^{2}\left|\phi_{3}\right|^{2}+\left|\phi_{2}\right|^{2}\left|\phi_{3}\right|^{2}-\left|\phi_{1}\right|^{2}\left|\phi_{4}\right|^{2} \nonumber \\
&+\left|\phi_{2}\right|^{2}\left|\phi_{4}\right|^{2}-\frac{1}{2}\left|\phi_{3}\right|^{2}\left|\phi_{4}\right|^{2}+\frac{1}{4}\left|\phi_{3}\right|^{4}+\frac{1}{4}\left|\phi_{4}\right|^{4}. 
\end{align}
$\varepsilon^{\pm}_{\text{zig-zag}}$ displays linear dispersing bands that go to zero at $\Gamma $ and $M$ points. We note the two negative eigenvalues are ignored.

\bibliography{MOF_Refs}